\documentclass[usenatbib, longauth,letter]{aa} % for the letters
\usepackage{graphicx}
\usepackage{graphics}
\usepackage{txfonts}
\usepackage{epsfig}
\usepackage{natbib}
\usepackage{float}
\usepackage{caption}
\usepackage{hyperref}
\usepackage{dblfloatfix}
\usepackage[]{color}
\begin{document}

\title{Mass distribution in the Galactic Center based on interferometric astrometry of multiple stellar orbits}

\titlerunning{Interferometric Astrometry of Multiple Stellar Orbits}
\subtitle{}

\author{GRAVITY Collaboration\thanks{GRAVITY is developed
    in a collaboration by MPE, LESIA of Paris Observatory / CNRS / Sorbonne Universit\'e
    / Univ. Paris Diderot and IPAG of Universit\'e Grenoble Alpes /
    CNRS, MPIA, Univ. of
    Cologne, CENTRA - Centro de Astrofisica e Gravita\c c\~ao, and
    ESO. Corresponding authors: S.~Gillessen (ste@mpe.mpg.de), F.~Widmann (fwidmann@mpe.mpg.de), \& G.~Hei{\ss}el (gernot.heissel@obspm.fr).
    }:
R.~Abuter\inst{8}
\and N.~Aimar\inst{2}
\and A.~Amorim\inst{6,12}
\and J.~Ball\inst{17}
\and M.~Baub\"ock\inst{1,15}
\and J.P.~Berger\inst{5,8}
\and H.~Bonnet\inst{8}
\and G.~Bourdarot\inst{5,1}
\and W.~Brandner\inst{3}
\and V.~Cardoso\inst{12, 14}
\and Y.~Cl\'{e}net\inst{2}
\and Y.~Dallilar\inst{1}
\and R.~Davies\inst{1}
\and P.T.~de~Zeeuw\inst{10,1}
\and J.~Dexter\inst{13}
\and A.~Drescher\inst{1}
\and F.~Eisenhauer\inst{1}
\and N.M.~F\"orster~Schreiber\inst{1} 
\and A.~Foschi\inst{12,14}
\and P.~Garcia\inst{7,12}
\and F.~Gao\inst{16,1}
\and E.~Gendron\inst{2}
\and R.~Genzel\inst{1,11}
\and S.~Gillessen\inst{1}
\and M.~Habibi\inst{1}
\and X.~Haubois\inst{9}
\and G.~Hei{\ss}el\inst{2}
\and T.~Henning\inst{3}
\and S.~Hippler\inst{3}
\and M.~Horrobin\inst{4}
\and L.~Jochum\inst{9}
\and L.~Jocou\inst{5}
\and A.~Kaufer\inst{9}
\and P.~Kervella\inst{2}
\and S.~Lacour\inst{2}
\and V.~Lapeyr\`ere\inst{2}
\and J.-B.~Le~Bouquin\inst{5}
\and P.~L\'ena\inst{2}
\and D.~Lutz\inst{1}
\and T.~Ott\inst{1}
\and T.~Paumard\inst{2}
\and K.~Perraut\inst{5}
\and G.~Perrin\inst{2}
\and O.~Pfuhl\inst{8,1}
\and S.~Rabien\inst{1}
\and J.~Shangguan\inst{1}
\and T.~Shimizu\inst{1}
\and S.~Scheithauer\inst{3}
\and J.~Stadler\inst{1}
\and A.W.~Stephens\inst{17}
\and O.~Straub\inst{1}
\and C.~Straubmeier\inst{4}
\and E.~Sturm\inst{1}
\and L.J.~Tacconi\inst{1}
\and K.R.W.~Tristram\inst{9}
\and F.~Vincent\inst{2}
\and S.~von~Fellenberg\inst{1}
\and F.~Widmann\inst{1}
\and E.~Wieprecht\inst{1}
\and E.~Wiezorrek\inst{1} 
\and J.~Woillez\inst{8}
\and S.~Yazici\inst{1,4}
\and A.~Young\inst{1}
}

\institute{
Max Planck Institute for extraterrestrial Physics,
Giessenbachstra{\ss}e~1, 85748 Garching, Germany
\and LESIA, Observatoire de Paris, Universit\'e PSL, CNRS, Sorbonne Universit\'e, Universit\'e de Paris, 5 place Jules Janssen, 92195 Meudon, France
\and Max Planck Institute for Astronomy, K\"onigstuhl 17, 
69117 Heidelberg, Germany
\and $1^{\rm st}$ Institute of Physics, University of Cologne,
Z\"ulpicher Stra{\ss}e 77, 50937 Cologne, Germany
\and Univ. Grenoble Alpes, CNRS, IPAG, 38000 Grenoble, France
\and Universidade de Lisboa - Faculdade de Ci\^encias, Campo Grande,
1749-016 Lisboa, Portugal 
\and Faculdade de Engenharia, Universidade do Porto, rua Dr. Roberto
Frias, 4200-465 Porto, Portugal 
\and European Southern Observatory, Karl-Schwarzschild-Stra{\ss}e 2, 85748
Garching, Germany
\and European Southern Observatory, Casilla 19001, Santiago 19, Chile
\and Sterrewacht Leiden, Leiden University, Postbus 9513, 2300 RA
Leiden, The Netherlands
\and Departments of Physics and Astronomy, Le Conte Hall, University
of California, Berkeley, CA 94720, USA
\and CENTRA - Centro de Astrof\'{\i}sica e
Gravita\c c\~ao, IST, Universidade de Lisboa, 1049-001 Lisboa,
Portugal
\and Department of Astrophysical \& Planetary Sciences, JILA, Duane Physics Bldg., 2000 Colorado Ave, University of Colorado, Boulder, CO 80309, USA
\and CERN, 1 Esplanade des Particules, Gen\`eve 23, CH-1211, Switzerland
\and Department of Physics, University of Illinois, 1110 West Green Street, Urbana, IL 61801, USA
\and Hamburger Sternwarte, Universit\"at Hamburg, Gojenbergsweg 112, 21029 Hamburg, Germany
\and Gemini Observatory/NSF's- NOIRLab, 670 N. A'ohoku Place, Hilo, Hawaii, 96720, USA
}

\date{Draft version \today}
% \abstract{}{}{}{}{}  5 {} token are mandatory  \abstract
%  % context heading (optional)
%  % {} leave it empty if necessary     {}
%  % aims heading (mandatory)   {t.b.d.}
%  % methods heading (mandatory)   {t.b.d.}
%  % results heading (mandatory) {t.b.d.}
%  % conclusions heading (optional), leave it empty if necessary    {}

\abstract{Stars orbiting the compact radio source Sgr~A* in the Galactic Center serve as precision probes of the gravitational field around the closest massive black hole. In addition to adaptive optics-assisted astrometry (with NACO / VLT) and spectroscopy (with SINFONI / VLT, NIRC2 / Keck and GNIRS / Gemini) over  three decades, we have obtained 30-100$\,\mu$as astrometry since 2017  with the four-telescope interferometric beam combiner GRAVITY / VLTI, capable of reaching a sensitivity of $m_K = 20$ when combining data from one night. 
We present the simultaneous detection of several stars within the diffraction limit of a single telescope, illustrating the power of interferometry in the field. The new data for the stars S2, S29, S38, and S55 yield significant accelerations between March and July 2021, as these stars pass the pericenters of their orbits between 2018 and 2023. This allows for a high-precision determination of the gravitational potential around Sgr~A*. Our data are in excellent agreement with general relativity orbits around a single central point mass, $M_\bullet=4.30\times10^6 M_\odot$, with a precision of about $\pm 0.25\%$. We improve the significance of our  detection of the Schwarzschild precession in the S2 orbit to $7\sigma$.
Assuming plausible density profiles, the extended mass component inside the S2 apocenter ($\approx 0.23''$ or $2.4\times10^4 R_\mathrm{S}$) must be $\lesssim 3000 M_\odot\, (1\sigma)$, or $ \lesssim 0.1$\% of $M_\bullet$. Adding the enclosed mass determinations from 13 stars orbiting Sgr~A* at larger radii, the innermost radius at which the excess mass beyond Sgr~A*  is tentatively seen is  $r \approx 2.5'' \geq 10\times$ the apocenter of S2. This is in full harmony with the stellar mass distribution (including stellar-mass black holes) obtained from the spatially resolved luminosity function. }


\maketitle

\section{Introduction}
\label{sec:intro}

The GRAVITY instrument on the Very Large Telescope Interferometer (VLTI) has made it possible to monitor the positions of stars within 0.1'' from Sgr~A*, the massive black hole (MBH) at the Galactic Center (GC), with a precision of $\approx 50\mu$as \citep{2017A&A...602A..94G}. The GRAVITY data taken in 2017-2019 together with the adaptive optics (AO) and Speckle data sets obtained since 1992 (at ESO telescopes), or since 1995 at the Keck telescopes have delivered exquisite coverage of the 16-year highly elliptical orbit of the star S2, which passed its most recent pericenter in May 2018. Besides the direct determinations of the mass of Sgr~A* ($M_\bullet$) and the distance to the GC ($R_0$), these interferometric data have provided strong evidence for general relativistic effects caused by the central MBH on the orbit of S2, namely, the gravitational redshift and the prograde relativistic precession \citep{2018A&A...615L..15G, 2019A&A...625L..10G, 2020A&A...636L...5G, 2021A&A...647A..59G, 2019Sci...365..664D}.

Due to its short period and  brightness, S2 is the most prominent star in the GC, but ever higher quality, high-resolution imaging and spectroscopy of the nuclear star cluster over almost three decades have delivered orbit determinations for some 50 stars \citep{2002Natur.419..694S, 2003ApJ...596.1015S, 2003ApJ...586L.127G, 2008ApJ...689.1044G, 2005ApJ...628..246E, 2009ApJ...692.1075G, 2017ApJ...837...30G, 2012Sci...338...84M, 2016ApJ...830...17B}. The motions of these stars show that the gravitational potential is dominated by a compact, central mass of $\approx 4.3\times10^6 M_\odot$ that is concentrated within S2's (3D) pericenter distance of $14\,$mas (or $120\,$AU) and 1400 times the event horizon radius $R_\mathrm{S}$ of a Schwarzschild (non-rotating) MBH for a distance of $8.28\,$kpc \citep{2019A&A...625L..10G, 2021A&A...647A..59G}.

S2 passes its pericenter with a mildly relativistic orbital speed of $7700\,$km/s ($\beta=v/c=0.026$). Based on the monitoring of the star's radial velocity and motion on the sky from data taken prior to and up to two months after pericenter, \cite{2018A&A...615L..15G} were able to detect the first post-Newtonian effects of general relativity (GR), namely:\ the gravitational redshift, along with the transverse Doppler effect of special relativity. The combined effect for S2 shows up as a $200\,$km/s residual centered on the pericenter time, relative to the Keplerian orbit with the same parameters. \cite{2019A&A...625L..10G} improved the statistical robustness of the detection of the gravitational redshift to $20\sigma$. \cite{2019Sci...365..664D} confirmed these findings from a second, independent data set mainly from the Keck telescope. While the redshift occurs solely in the wavelength space, the superior astrometry of GRAVITY sets much tighter constraints on the orbital geometry, mass, and distance, all the while decreasing the uncertainty on the redshift parameter more than three times relative to data sets from single telescopes. 

The precession due to the Schwarzschild metric is predicted to lead to a prograde rotation of the orbital ellipse in its plane of $\Delta \omega = 12.1'$ per revolution for S2,  corresponding to a shift in the milli-arcsec regime of the orbital trace on sky; hence 
using interferometry is particularly advantageous in this case.
\cite{2020A&A...636L...5G} detected the Schwarzschild precession at the $5\sigma$ level. The uncertainties on the amount of precession can then be interpreted as limits on how much extended mass (leading to retrograde precession) might be present within the S2 orbit.

Here, we expand our analysis by two more years, up to 2021.6. We combine GRAVITY data from four stars, alongside with the previous AO data. Section~\ref{sec:observations} presents the new data and Section~\ref{sec:analysis} describes our analysis. In Section~\ref{sec:sp}, we show the combined fits, improving the accuracy of the measured post-Newtonian parameters of the central black hole and the limits on the extended mass (Section~\ref{sec:limit}). In combination with earlier measurements of stars with larger apocenters, we study the  mass distribution out to $\approx$ 3''. Section~\ref{sec:conclusions} summarizes our conclusions.

\begin{figure*}[t!]
\centering
\includegraphics[width=10.9cm]{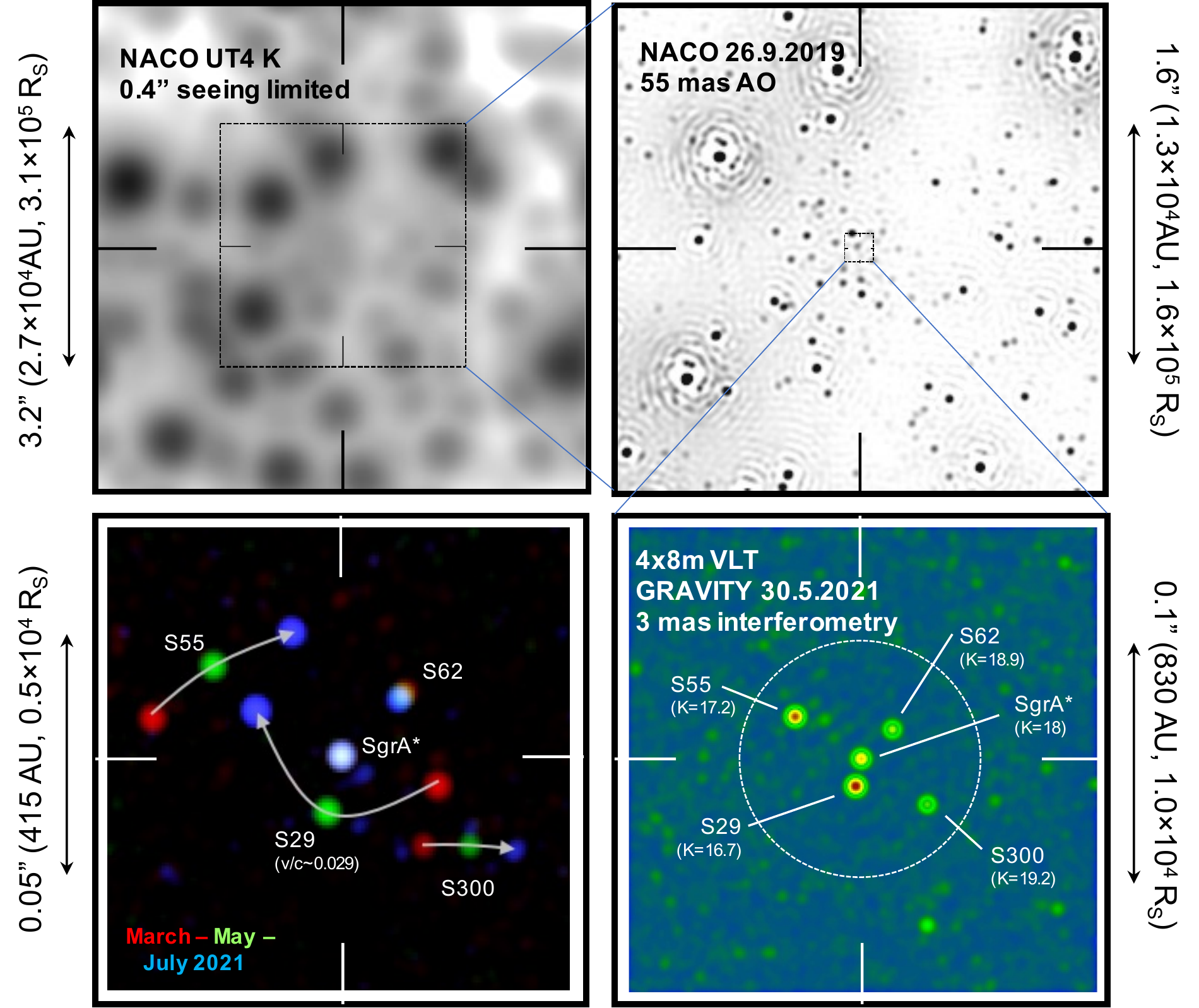}
\caption{Progress in stellar astrometric imaging in the GC, from seeing limited observations ($\approx 0.4-0.5''$ FWHM, top left) to AO imaging on $8-10\,$m class telescopes ($50-60\,$mas FWHM, $\approx 300-500\,\mu$as astrometry (top-right), showing a Lucy-Richardson deconvolved and beam-restored AO image), to current interferometric imagery with GRAVITY / VLTI (bottom panels, $2\times4$ mas FWHM resolution, $30-100\,\mu$as astrometry). Bottom right: Central $0.2''$ region centered on Sgr~A*(IR) on May 30, 2021 (see Appendix~\ref{sec:A2}, \citealt{JULIA}). Over 2021, the central GRAVITY field of view was dominated by four stars (S29, S55, S62, and S300) with K-band magnitudes between 16.7 and 19.2, in addition to variable emission from Sgr~A*(IR) itself. The dashed circle indicates the field-of-view of GRAVITY, defined by the Gaussian acceptance profile of the pick-up fibres in the instrument. Comparing in the bottom left panel three such images taken on March 30 (red), May 30 (green), and July 26 (blue), the orbital motions of all four stars are easily seen, topped by the $\approx8740\,$km/s  velocity of S29 at its pericenter on 2021.41 ($R_\mathrm{peri} \approx 100\,$AU). }
\label{fig1}
\end{figure*}

\section{Observations}
\label{sec:observations}

Interferometric astrometry with GRAVITY has several distinct advantages over single-telescope, AO imaging (Figure~\ref{fig1}): 

First, the higher angular resolution yields an order of magnitude better astrometric precision for isolated sources. 

Secondly, for crowded environments, such as the central arcsecond that has a surface density $> 100$ stars per square arcsecond to $K < 17$ (and more for fainter limits, \citealt{2003ApJ...594..812G, 2018A&A...609A..28B, 2018MNRAS.476.3600W}),  interferometric data  are much less affected (by a factor of several hundred) by confusion noise. In the context of the GC cluster imaging, this issue was recognized early on \citep{2003ApJ...586L.127G, 2008ApJ...689.1044G, 2017ApJ...837...30G, 2019Sci...365..664D, 2020A&A...636L...5G}. For "orbit crossings" of modest duration for individual brighter stars, this often means that data over a duration of a few years are affected. The situation is  worse at the pericenter passage of S2 (2002, 2018), when the star and the variable source Sgr~A* are in the same diffraction beam of an $8-10\,$m class telescope \citep{2008ApJ...689.1044G}. For 2021/2022, our data show explicitly that in addition to Sgr~A*, several stars are present in the central beam (see also \citealt{JULIA}), making single-telescope astrometry even more uncertain or unusable. 

Third, close to Sgr~A*, astrometry with interferometry reduces to fitting the phases and visibilities with a multiple point source model in a single pointing of the interferometric fiber, which is straightforward, once the optical aberrations across the fiber field of view are corrected for \citep{2020A&A...636L...5G, 2021A&A...647A..59G}. Interferometric measurements beyond the fibre field of view require double pointings, which, thanks to GRAVITY's metrology system, can be related astrometrically to each other (Appendix~\ref{sec:A12}).
The GRAVITY positions are directly referring to Sgr~A*, since it is visible in each exposure and since it is one of the point sources in the multi-source model. In contrast, AO astrometry relies on establishing a global reference frame by means of stars visible 
both at radio wavelengths (where Sgr~A* is visible as well) and in the near-infrared. A few such SiO maser stars exist in the GC \citep{2007ApJ...659..378R} and they reside at larger separations ($\approx 20''$). Hence, to establish an AO-based reference frame, it is necessary to correct for the distortions of the imaging system \citep{2015MNRAS.453.3234P, 2019ApJ...873...65S, 2019ApJ...873....9J}.

In addition to S2 (now moving away from Sgr~A* since its 2018.38 pericenter), we used the stars S29 ($K=17.6$, pericenter 2021.41), S38 ($K=18.3$, 2022.70), and S55 ($K=17.2$, 2021.7). The latter star was labeled S0-102 in \cite{2012Sci...338...84M}. These authors determined its orbital period to be 12 years.

\begin{itemize}
\item For S2, we included 128 NACO astrometric data points, 92 SINFONI, 3 Keck (2000-2002), 2 NACO (2003) and 4 GNIRS / GEMINI spectra, and 82 GRAVITY astrometric measurements. Compared to \cite{2020A&A...636L...5G}, we added the 4  GNIRS spectra, and 17 GRAVITY epochs. These data cover the time span of 1992.2-2021.6.
\item For S29, we included 94 NACO, 17 SINFONI, 2 GNIRS, and 21 GRAVITY measurements, covering 2002.3-2021.6.
\item For S38, we included 110 NACO, 8 SINFONI, 1 Keck, and 6 GRAVITY measurements, covering 2004.2-2021.6.
\item For S55, we included 42 NACO, 2 SINFONI, and 18 GRAVITY measurements, covering 2004.5-2021.6.
\item We also analyzed the NACO and SINFONI 2002-2019 data for another 13 stars with $K\approx 12-16$ that have sufficient data to infer a mass (and their orbital parameters), with $R_0$
fixed to $8277\,$pc (the best fit value, Sec.~\ref{sec:sp}). 
\end{itemize}
We report the full list of observations in Appendix~\ref{tab:obslist}.

\section{Analysis}
\label{sec:analysis}

For a single-star fit, we typically fit for 14 parameters: six parameters describing the initial osculating Kepler orbit $(a, e, i, \omega, \Omega, t_0)$, $R_0$ and $M_\bullet$, along with five coordinates describing the on-sky position and the 3D velocity of the mass (relative to the AO spectroscopic/imaging frame), and a dimensionless parameter encoding the non-Keplerian effect we are testing for. For the gravitational redshift, we used $f_\mathrm{gr}$, which is 0 for Newtonian orbits and 1 for GR-orbits. In \cite{2018A&A...615L..15G} we found 
$f_\mathrm{gr} = 0.90 \pm 0.17$; and in \cite{2019A&A...625L..10G}, we found $f_\mathrm{gr}=1.04\pm0.05$. \cite{2019Sci...365..664D} reported $f_\mathrm{gr} = 0.88 \pm 0.17$. 

For the Schwarzschild precession, we use the first-order post-Newtonian expansion for a massless test particle \citep{1993tegp.book.....W} and add a factor $f_\mathrm{SP}$ in the equation of motion in front of the 1PN terms, where $f_\mathrm{SP}=0$ corresponds to Keplerian motion and  $f_\mathrm{SP}=1$ to GR. In \cite{2020A&A...636L...5G}, we found $f_\mathrm{SP}=1.10\pm0.19$. 

Similarly, we parameterize an extended mass distribution by including a parameter $f_\mathrm{Pl}$ in the normalization of the profile.  Following \cite{2017ApJ...837...30G} and \cite{2020A&A...636L...5G}, we assume a  \cite{1911MNRAS..71..460P} profile
\begin{equation}
\rho(r) = \frac{3 f_\mathrm{Pl} M_\bullet}{4 \pi\, a_\mathrm{Pl}^3} \times \left(1+\left(\frac{R}{a_\mathrm{Pl}}\right)^2\right)^{-5/2} \,\,
,\end{equation}
with scale length, $a_\mathrm{Pl}$, and total mass of $f_\mathrm{Pl} M_\bullet$. 
We use  $a_\mathrm{Pl}= 1.27 a_\mathrm{apo}(\mathrm{S2})=0.3''$ \citep{2005AN....326...83M}. The enclosed mass within $R$ is $M(\le R)  =  f_\mathrm{Pl} M_\bullet \,(R/a_\mathrm{Pl})^3 \,(1+R^2/a^2_\mathrm{Pl} )^{-3/2}$. We fit for the fraction of $M_\bullet$ that is in the extended mass, $f_\mathrm{Pl}$.

Following \cite{2018A&A...615L..15G, 2019A&A...625L..10G, 2020A&A...636L...5G}, we find the best-fit values
by fitting simultaneously all parameters, including prior constraints. Throughout our study, we used an outlier robust fitting (Sect 3.2 in \citealt{2020A&A...636L...5G}). The inferred uncertainties are affected and partially dominated by systematics, especially when combining data from three or more measurement techniques. Our parameterization keeps correlations between $f_\mathrm{SP}$ or $f_\mathrm{Pl}$ and $M_\bullet$ or $R_0$ small, but the two parameters of interest show some degeneracy with the coordinate system offsets. 

To check the formal fit errors, we carried out a (Metropolis-Hastings) Markov-Chain Monte Carlo (MCMC) analysis. Using 100~000 realizations, we found the distributions and parameter correlations of the respective dimensionless parameter, $f_\mathrm{SP}$ or $f_\mathrm{Pl}$, with the other parameters and test whether they are well described by Gaussian distributions. 
For more details, see \cite{2018A&A...615L..15G, 2019A&A...625L..10G, 2020A&A...636L...5G, 2021A&A...647A..59G} and Appendix~\ref{sec:A}.

\begin{figure*}[t!]
\centering
%\vspace{-0.7cm}
\includegraphics[width=13.8cm]{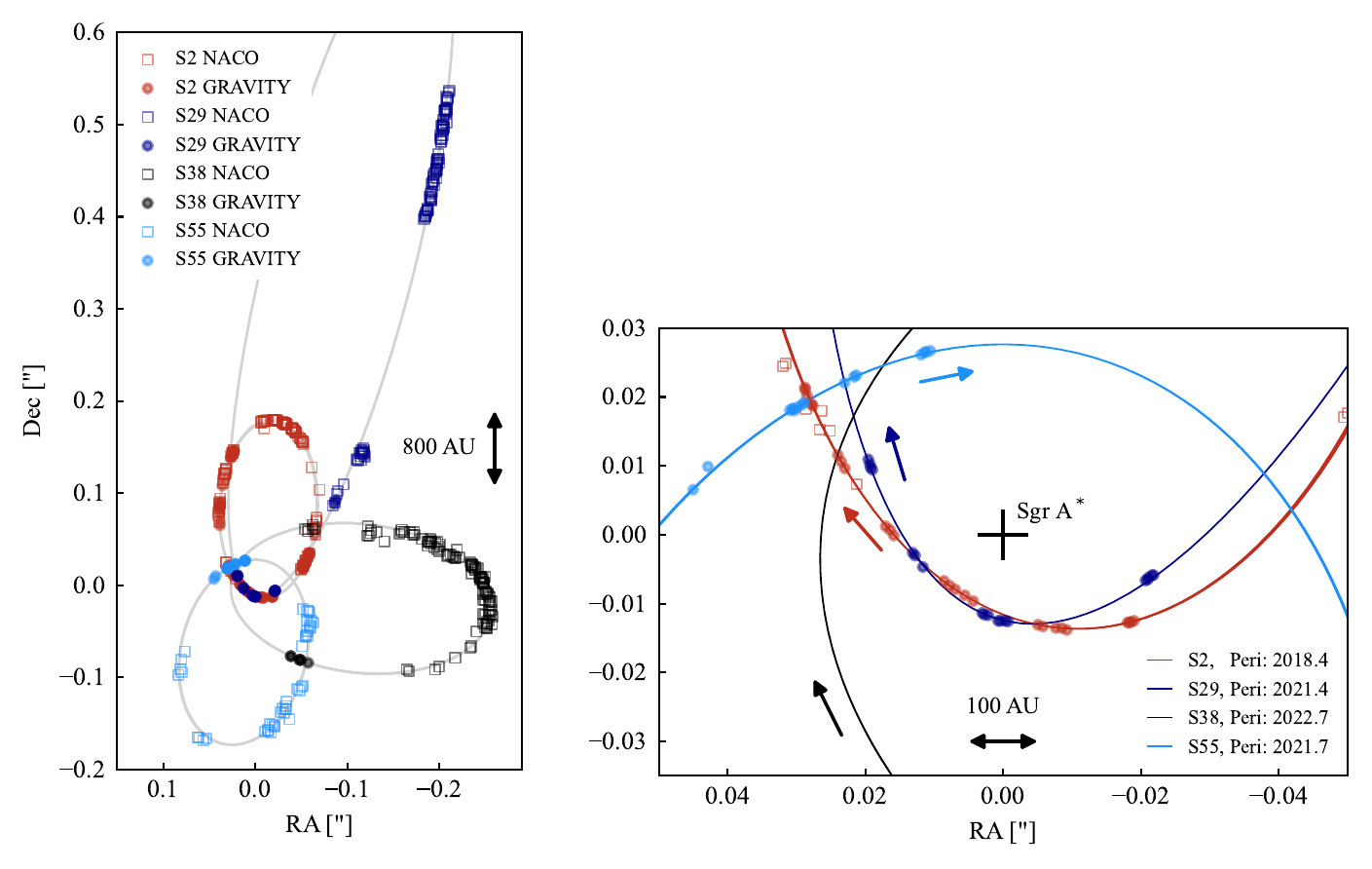}
%\vspace{-3cm}
\caption{Summary of the 1992-2021 astrometric data and best fitting orbits for the stars S2, S29, S38, and S55. The black cross marks the position of the compact radio source Sgr~A* and mass center. The directions of projected orbital motions are marked by arrows.}
\label{fig2}
\end{figure*}

In \cite{2009ApJ...692.1075G, 2017ApJ...837...30G}, we consistently found, based on the AO data, that the basic parameters describing the gravitational potential ($M_\bullet$ and $R_0$) and the extended mass (e.g., $f_\mathrm{Pl}$ for an assumed Plummer distribution) are best constrained by the S2 data. Including other stars only moderately improved the fitting quality and uncertainties. This is because of the superior number and quality of the S2 data compared to those of the other stars. Since the higher resolution GRAVITY astrometry became available, S2 has completely dominated our knowledge about the central potential. Another reason is that it is only for S2 that we have data at or near pericenter, which are the most sensitive component of the data to the mass distribution, as the explicit analysis in \cite{2017ApJ...837...30G} and \cite{HEISSEL} shows. 

This situation changes with the data set used here. We now have GRAVITY data of four stars with comparable pericenters at $12\,$mas (S29), $14\,$mas (S2), $26\,$mas (S38), and $29\,$mas (S55). Naturally, we need to fit for $4 \times 6$ orbital parameters $(a_i, e_i, i_i, \omega_i, \Omega_i, t_{0,i})$ in addition to the NACO/SINFONI zero points $(x_0, y_0, vx_0, vy_0, vz_0)$, $M_\bullet$ and $R_0$, as well as $f_\mathrm{SP}$ and/or $f_\mathrm{Pl}$. However, the inclusion of near-pericenter GRAVITY data of S29, S38, and S55 lessens the parameter correlations and uncertainties (Figure~\ref{fig2}); this is also because the orbits are oriented almost perpendicular to each other in at least one of the Euler angles. 
Furthermore, the orbits of the four stars probe the precession over a wider parameter range in  semi-major axis and eccentricity.

A comparison of the results obtained with different fitting schemes and codes of the consortium (of MPE, Univ. Cologne and LESIA) underlines the importance of two aspects of the data analysis: First, AO-based astrometry of stars near pericenter is subject to confusion, with the emission from Sgr~A* and other neighboring stars contributing to the centroid. These data now can be discarded in favor of better-resolution interferometric data. Second, the NACO-frame zero point and drift on the one hand, and the pro- or retrograde precession on the other hand, are degenerate. To reduce this degeneracy, we can use three sources of additional information: (i) NACO astrometry of flares, (ii) the prior from the construction of the AO reference frame \citep{2015MNRAS.453.3234P}, or (iii) the orbits of further stars that have sufficient phase coverage to constrain the zero points. In \cite{2020A&A...636L...5G}, we combined (i) and (ii). Avoiding the potentially confused flare positions (i), here we use the combination of (ii) and the orbits of 13 further stars (iii). 
We derive $x_0=0.57\pm0.15\,$mas, $y_0=-0.06\pm0.25\,$mas (both epoch 2010.35), $vx_0 =63 \pm 7 \,\mu$as/yr, $vy_0=33 \pm 2\,\mu$as/yr, which are consistent with the earlier estimates, but with smaller uncertainties.

\section{Schwarzschild precession for S2}
\label{sec:sp}
\begin{figure*}[t!]
\centering
%\vspace{-0.7cm}
\includegraphics[width=16cm]{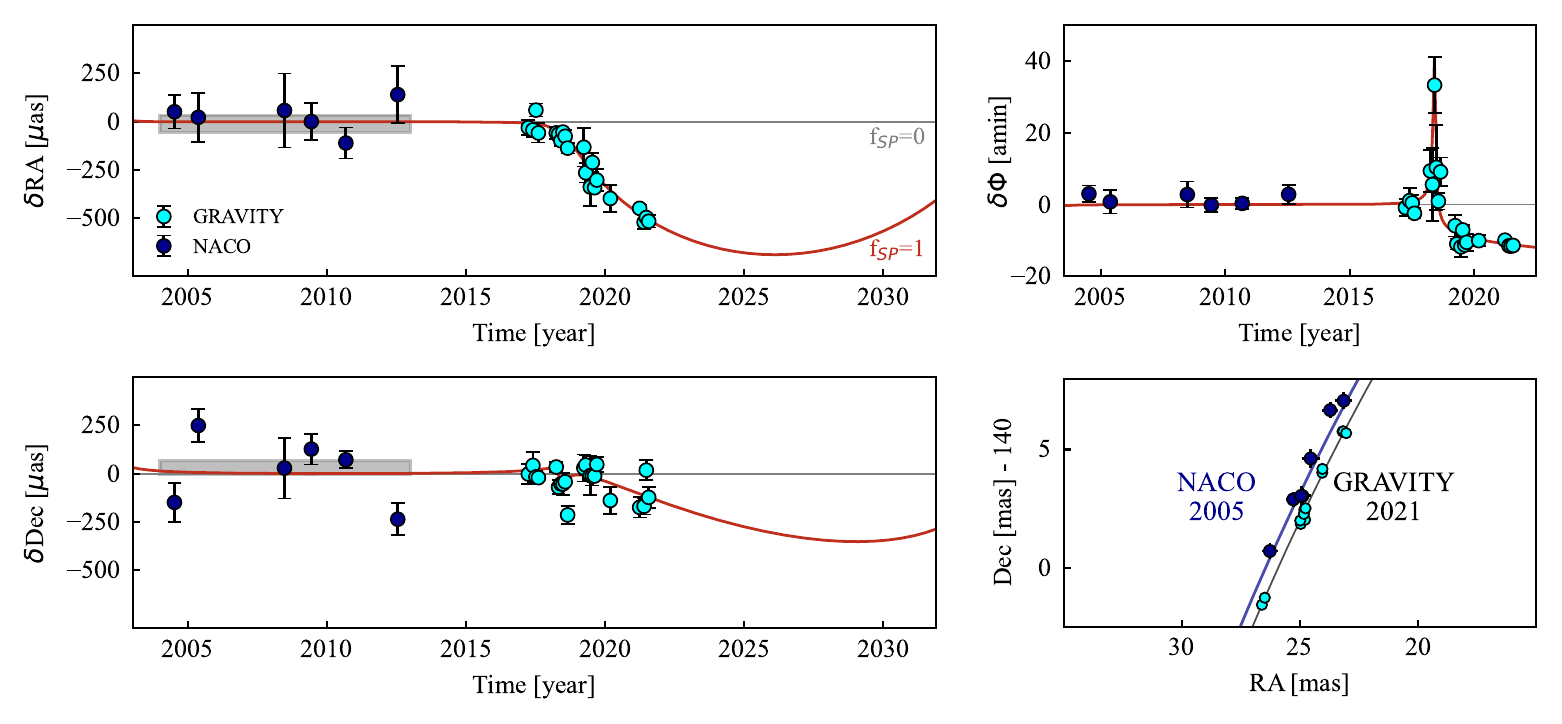}
%\vspace{-3cm}
\caption{Illustration of the fit results for $f_\mathrm{SP}$. We show the key figures for S2. The two left panels show the residuals in RA (top) and Dec (bottom) between the GRAVITY (cyan filled circles, and  $1\sigma$ uncertainties) and NACO data (blue filled circles) and the best GR fit (red curve, $f_\mathrm{SP}=1$, R{\o}mer effect, plus special relativity, plus gravitational redshift and Schwarzschild precession), relative to the same orbit for $f_\mathrm{SP}=0$ (Kepler/Newton, plus R{\o}mer effect, plus special relativity, plus redshift). The orbital elements for non-Keplerian orbits (i.e., with $f_\mathrm{SP} \ne 0$) are interpreted as osculating parameters at apocenter time, 2010.35. NACO and GRAVITY data are averages of several epochs. The grey bars denote averages over all NACO data near apocenter (2004-2013). Top right: the same for the residual orbital angle on the sky $\delta \phi =\phi(f_\mathrm{SP}=1)-\phi(f_\mathrm{SP}=0)$. Bottom right: Zoom into the 2005/2021 part of the orbit, plotted in the mass rest frame. The earlier orbital trace does not coincide with the current one due to the Schwarzschild precession.}
\label{fig4}
\end{figure*}

Repeating the analysis of \cite{2020A&A...636L...5G} (S2 alone but with the updated zero points) and solving for the Schwarzschild precession parameter we find $f_\mathrm{SP}=0.85\pm0.16$ ($\chi^2_r=1.11$). This is naturally very similar to our previous results\footnote{Following strictly \cite{2020A&A...636L...5G}, i.e., using the S2 data plus the flare positions of Sgr~A* with the zero point priors of \cite{2015MNRAS.453.3234P} yields $f_\mathrm{SP}=1.23\pm0.14$ ($\chi^2_r=1.70$).}, but the new data have decreased the $1\sigma$ uncertainty from $\pm0.19$ to $\pm0.16$.

Next, we fit with the four star (S2, S29, S38, S55) data, and find $f_\mathrm{SP}=0.997 \pm0.144$, with $\chi^2_r=2.17$, (Figure~\ref{fig2}). Figure~\ref{fig4} shows the residuals of the best fit and from the corresponding Newtonian ($f_\mathrm{SP}=0$) orbit. The combination of the near-pericenter GRAVITY data of four stars improves the constraints on the common parameters. The contributions raising $\chi^2_r>1$ come from the NACO data of S29, S38, and S55 covering the outer parts of their orbits more affected by confusion.  

Applying MCMC analysis we find the most likely values of $f_\mathrm{SP}=0.85\pm0.18$ (only S2) and $f_\mathrm{SP}=0.99\pm0.15$ (S2, S29, S38, S55). Figure~\ref{figApp1} shows the full set of parameter correlations, including the well-known degeneracy between $M_\bullet$ and $R_0$ \citep{2008ApJ...689.1044G, 2016ApJ...830...17B, 2009ApJ...692.1075G, 2017ApJ...837...30G}. All of the 32 parameters of the four-star fit are well constrained.

As discussed by \cite{2020A&A...636L...5G}, the impact of the high eccentricity of the S2 orbit ($e=0.88$) is that most of the precession happens in a short time-frame around pericenter. Due to the geometry of the orbit most of the precession shows up in the RA-coordinate, and the change in $\omega$ after pericenter appears as a kink in the RA-residuals. The data are obviously in excellent agreement with GR. Compared to \cite{2020A&A...636L...5G}, the significance of this agreement has improved from 5 to $7\,\sigma$, from the combination of adding two more years of GRAVITY data to the S2 data set and the expansion to a four-star fit. Table~\ref{tab1} gives the best fit orbit parameters, zero points, $M_\bullet$, and $R_0$.

As of 2021, S2 is sufficiently far away from pericenter, such that the Schwarzschild precession can now be seen as a $\approx0.6\,$mas shift between the data sets in RA (and less so in Dec) between two consecutive passages of the star on the apocenter-side of the orbit. This effect is obvious when comparing the 2021 GRAVITY data to the 2005 NACO data, exactly one period prior (Figure~\ref{fig4} right). This comparison illustrates that the Schwarzschild precession dominates the entire orbit and that there is no detectable retrograde (Newtonian) precession due to an extended mass component (see \citealt{HEISSEL}).

\section{Limits on extended mass}
\label{sec:limit}
\begin{figure}[t!]
\centering
%\vspace{-0.7cm}
\includegraphics[width=8.8cm]{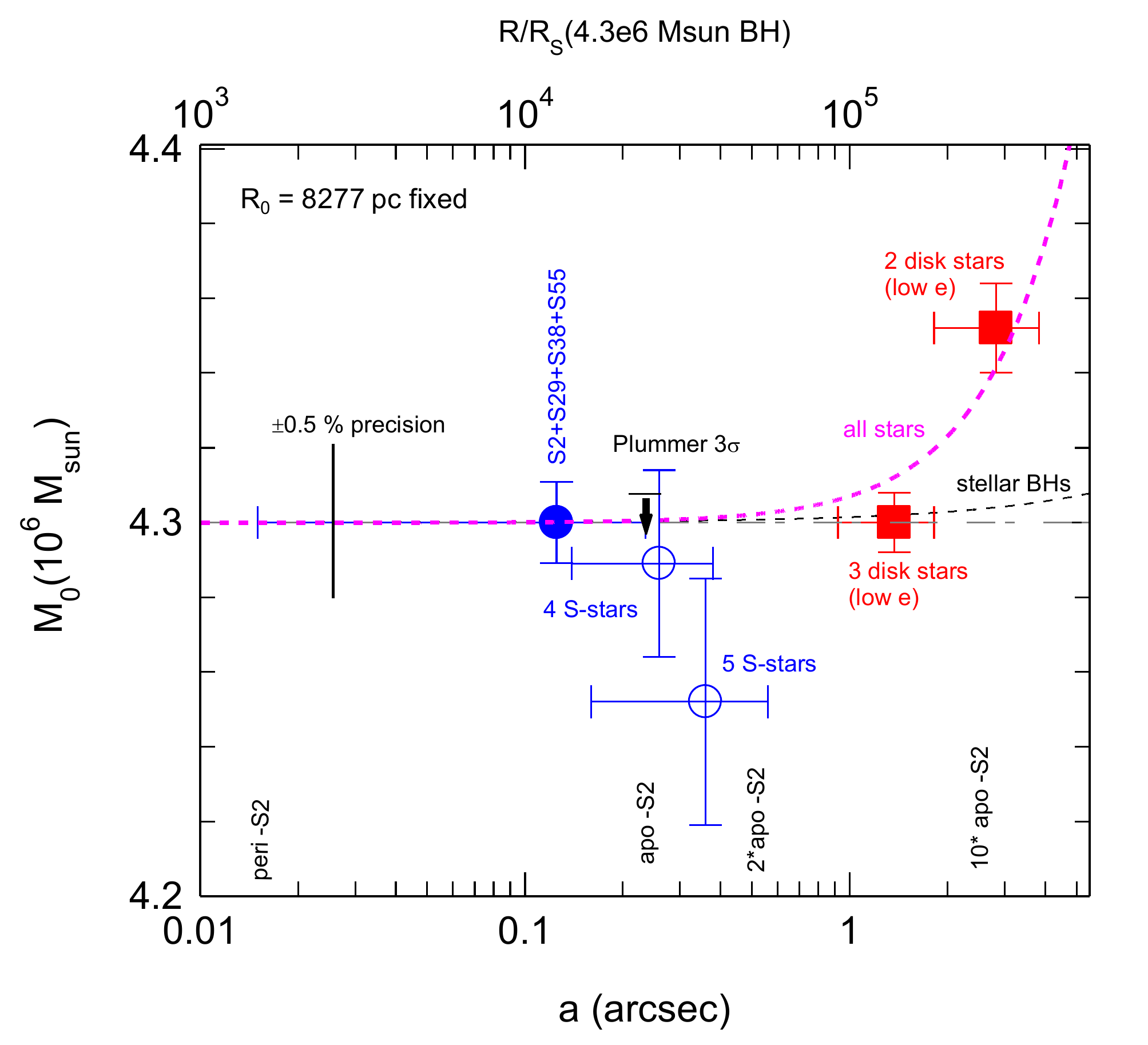}
%\vspace{-3cm}
\caption{Summary of the central mass distribution in the GC, within 20$\times$ the apocenter radius of the S2 orbit, $\approx 5''$. The filled blue circle is the central mass of $4.30\times10^6 M_\odot$ (for $R_0=8277\,$pc, and its $1 \sigma$ uncertainty of $12,000 M_\odot$), which our four-star fitting  has established to lie within the $1200 R_\mathrm{S}$ pericenter of S29. The black arrow denotes the $3\sigma$ upper limit of $M_\mathrm{ext}(\le0.3'')$ of the sum of $M_\bullet$ and any extended Plummer mass of assumed scale radius 0.3''. The two open blue circles and two red filled squares show averages of enclosed masses within the semi-major axes of other S-stars and clock-wise disk stars. 
The magenta dashed line is the sum of $M_\bullet$ and the extended stellar mass distribution from the literature (e.g., \citealt{2010RvMP...82.3121G, 2017ARA&A..55...17A, 2018A&A...609A..27S, 2018A&A...609A..28B}). All data are in excellent agreement with  a point mass (the MBH Sgr~A*) and a  star cluster with a power-law density slope $\gamma\approx1.6$ \citep{2003ApJ...594..812G, 2018A&A...609A..26G}, consisting of main-sequence stars and a small contribution of giants. No extra component from dark matter or an intermediate mass black hole $>10^3 M_\odot$ \citep{2020A&A...636L...5G} is required or compatible with the data.}
\label{fig6}
\end{figure}

In the following, we fix $f_\mathrm{SP}=1$ at its GR value and allow now for an extended mass component parameterized by $f_\mathrm{Pl}$.
We find $f_\mathrm{Pl}=(2.7 \pm 3.5) \times 10^{-3}$ from a single S2 fit and $f_\mathrm{Pl}=(-3.8 \pm 2.4) \times 10^{-3}$ for the four-star fit, in accordance with the findings of \cite{HEISSEL}. The latter $1\sigma$ error is consistent, albeit three to four times smaller than that of \cite{2017ApJ...837...30G} and 1.7 times smaller than that of \cite{2020A&A...636L...5G}, corresponding to $2400\,M_\odot$ within the apocenter of S2. 
In Figure~\ref{fig6}, we include the $3 \sigma$ uncertainty as a conservative upper limit, indicating that the extended mass cannot exceed $7500 M_\odot$. As in  \cite{2017ApJ...837...30G} and \cite{2020A&A...636L...5G}, we find again that varying $a_\mathrm{Pl}$ or replacing the Plummer distribution by a suitable power law changes this result trivially. Furthermore, we get a weaker limit by a factor of $\approx 2$ when omitting the NACO astrometry, using only GRAVITY \& SINFONI data.

The impact of an extended mass is naturally largest near apocenter of an orbit 
(e.g. \citealt{HEISSEL}). Figure~\ref{fig5} shows the result of adding various amounts of extended mass on top of the best-fit residuals with a point mass only. Our data are just commensurate with an additional $f_\mathrm{Pl} = 0.25\%$ of  $M_\bullet$, but a larger mass is excluded by both the near-peri- and near-apocenter data. The apparent sensitivity of the near-pericenter data in Figure~\ref{fig5} is the result 
of referring the residuals to the osculating Keplerian orbit at apocenter in 2010.35, such that the accumulating retrograde precession enters the near-pericenter data.

A second, independent measurement of the dynamically inferred mass distribution comes from fitting for the central mass using 13 individual stellar orbits with $a=0.1''$ to $3.8''$ \citep{2017ApJ...837...30G}, with $R_0$ and zero-points fixed to the best fitting values of the four-star fit (Figure~\ref{fig-orbits}). We then averaged the results in four groups of four stars with $0.11'' < a <0.22''$, five stars with $0.27'' < a < 0.4''$, three stars with $0.55'' < a <1.6''$, and two stars with $1.6''<a<3.8''$. Most of the stars in the first two groups are classical "S-stars" (mostly early-type B stars) with typically large eccentricities (e.g., \citealt{2008ApJ...689.1044G, 2017ApJ...837...30G}), while most of the stars in the outer groups are O and B stars in the clock-wise disk \citep{2006ApJ...643.1011P, 2009ApJ...697.1741B, 2009ApJ...690.1463L}, with modest eccentricities. The stars in the first two groups indicate that the mass is consistent with  $M_\bullet$ to within 0.3-0.6\%. There is no indication for an extended mass larger than $\approx25,000 M_\odot$ within $2\, a_\mathrm{apo}(\mathrm{S2})\approx 0.5''$. The outer stars suggests an extended mass of $15,000 M_\odot$ (and conservatively a $3\sigma$ limit of $50,000 M_\odot$) within $5-10 \,a_\mathrm{apo} (\mathrm{S2})$. 

Figure~\ref{fig6} summarizes the mass distribution within $5''$ ($\approx 20 \times$ the apocenter of S2). These estimates and limits are in excellent agreement with the distribution of stars (and stellar mass black holes and neutron stars) contained in this inner region around Sgr~A*, as estimated from models and simulations \citep{2017ARA&A..55...17A, 2018A&A...609A..28B}, or from observations of faint stars and diffuse stellar light (Figure~\ref{fig6}, \citealt{2010RvMP...82.3121G, 2018A&A...609A..26G, 2018A&A...609A..27S, 2019ApJ...872L..15H}).

In summary, several precise (O(0.1-0.3\%), $1\sigma$) determinations show that the mass distribution in the GC within 5'' $\approx 5\times10^5 R_\mathrm{S}$ of Sgr~A* is dominated by a central, compact mass. This mass is definitely enclosed within the pericenter of S29 ($12\,$mas, $\approx1200 R_\mathrm{S}$). Taking the gas motions at $\approx 3-5 \,R_\mathrm{S}$ \citep{2018A&A...618L..10G} and the mm-size of Sgr~A* \citep{2008Natur.455...78D, 2017ApJ...850..172J, 2019ApJ...871...30I} into account, the data are in excellent agreement with the MBH paradigm.

\section{Conclusions}
\label{sec:conclusions}

Here, we  present GRAVITY data obtained at the VLTI  in 2021. Within the central $20\,$mas, we observe the motions of four stars between March and July, illustrating the power of the interferometric, high spatial resolution. Using the novel astrometry of the stars S2, S29, S38, and S55,
along with new radial velocities obtained with GNIRS, we update our orbital analysis.

The star S2 has now returned to the part of its 16-year orbit for which good NACO AO-assisted positions were obtained during its previous passage. A direct comparison of the positions confirms that the orientation of the orbital ellipse has indeed shifted in its plane by the 12.1' expected from the prograde Schwarzschild precession induced by the gravitational field of the MBH, as reported in \cite{2020A&A...636L...5G}. 

At $K=14.1$, S2 is comparably bright. With its increased distance from Sgr~A* in 2021, we are now able to map with GRAVITY the immediate vicinity of the MBH to significantly fainter objects. This provided accurate positions for S29, S38, and S55.
These stars have previously measured NACO positions when they were further away from Sgr~A*. Combining these with the GRAVITY data improves the orbital parameters of the three stars substantially. In particular, S29 is on a deeply plunging ($e=0.97$) orbit with a period of $\approx\,$90 years and pericenter passage in 2021.41, with a space velocity of $\approx 8740\,$km/s at only $100\,$AU from Sgr~A*. 

S2, S29, S38, and S55 orbit in the same gravitational potential, and combining their astrometry and radial velocity data improves the accuracy of the determination of the properties of the central MBH. This leads to a 14\% measurement precision of the Schwarzschild precession, which is in full agreement with the prediction of GR. The best fit further yields $R_0 = (8277 \pm 9)\,$pc and $M_\bullet = (4.297 \pm 0.012) \times 10^6 M_\odot$  (statistical errors, see \citealt{2021A&A...647A..59G} for a discussion of the systematics that are  $\approx 30\,$pc for $R_0$ and $\approx 40,000 M_\odot$ for $M_\bullet$).

Any smooth extended mass distribution would lead to a retrograde precession of the S2 orbit relative to the relativistically precessing one and we can thus place a limit on a hypothetical mass distribution. The measurement errors leave room for at most $\lesssim 3000\,M_\odot$ in extended mass out to $230\,$mas. We included 13 stars further out with earlier measurements to trace the mass as a function of radius. The data are fully consistent with a single point mass, and only at $r\gtrsim2.5''$ does the enclosed mass tentatively exceed $M_\bullet$, which is consistent with the theoretically expected stellar mass distribution. Inside the $100\,$AU pericenter of S29, the orbits of Sgr~A* flares \citep{2018A&A...618L..10G}, together with the coincidence of mass location and light centroid \citep{2015MNRAS.453.3234P,2020ApJ...892...39R} constrain the mass distribution further, excluding, for example, dark matter spikes, as proposed by \cite{2020A&A...641A..34B, 2021MNRAS.505L..64B}, and the presence of an intermediate mass black hole as well \citep{2020A&A...636L...5G}.

Our multi-epoch GRAVITY data also confirm that at any time, there are likely a few stars that are sufficiently close to Sgr~A* on the sky to systematically influence its position derived with AO-assisted imaging on single telescopes. In addition, in 2022, two stars will pass the pericenters of their orbits at less than $100\,$mas distance (S38 and S42). The upgrade of GRAVITY to GRAVITY+ \citep{GPLUS} will push the sensitivity limit to $K>20$, which may  reveal more stars with even smaller orbits. The $39\,$m ELT equipped with MICADO \citep{2021Msngr.182...17D} and HARMONI \citep{2021Msngr.182....7T} might be the prime choice for obtaining radial velocities of such stars. Yet, GRAVITY+ will beat the ELT's angular resolution by a factor three, allowing continued $<50\,\mu$as astrometry and going even deeper than what we have demonstrated so far \citep{JULIA}.

\begin{acknowledgements}

We are very grateful to our funding agencies (MPG, ERC, CNRS [PNCG, PNGRAM], DFG, BMBF, Paris Observatory [CS, PhyFOG], Observatoire des Sciences de l'Univers de Grenoble, and the Funda\c c\~ao para a Ci\^encia e Tecnologia), to ESO and the Paranal staff, and to the many scientific and technical staff members in our institutions, who helped to make NACO, SINFONI, and GRAVITY a reality. S.G. acknowledges the support from ERC starting grant No. 306311. F.E. and O.P. acknowledge the support from ERC synergy grant No. 610058. A.A., A.F., P.G. and V.C. were supported by Funda\c{c}\~{a}o para a Ci\^{e}ncia e a Tecnologia, with grants reference 
SFRH/BSAB/142940/2018, UIDB/00099/2020 and PTDC/FIS-AST/7002/2020. Based on observations collected at the European Southern Observatory under the ESO programmes listed in Appendix~\ref{app:list}. The GNIRS spectra were obtained at the international Gemini Observatory, a program of NSF's NOIRLab, managed by the Association of Universities for Research in Astronomy (AURA) under a cooperative agreement with the National Science Foundation (NSF) on behalf of the Gemini Observatory partnership: the National Science Foundation (United States), National Research Council (Canada), Agencia Nacional de Investigaci\'{o}n y Desarrollo (Chile), Ministerio de Ciencia, Tecnolog\'{i}a e Innovaci\'{o}n (Argentina), Minist\'{e}rio da Ci\^{e}ncia, Tecnologia, Inova\c{c}\~{o}es e Comunica\c{c}\~{o}es (Brazil), and Korea Astronomy and Space Science Institute (Republic of Korea). This work was enabled by observations made from the Gemini North telescope, located within the Maunakea Science Reserve and adjacent to the summit of Maunakea. We are grateful for the privilege of observing the Universe from a place that is unique in both its astronomical quality and its cultural significance. We thank our anonymous referee for very useful comments.

\end{acknowledgements}

\bibliography{references}

\begin{appendix}

\section{Experimental techniques}
  \label{sec:A}
  
\subsection{GRAVITY: Determining astrometric separations}
\label{sec:A1}

The full width half maximum (FWHM) of the interferometric field of view (IFOV) of GRAVITY is $70\,\mathrm{mas}$. In consequence, not all stars discussed in this paper are observable simultaneously. The star S2 has moved too far away from Sgr~A* compared to 2018 to be observable simultaneously with Sgr~A*, while the stars S55 and S29 (and others, see \citealt{JULIA}) are always observed alongside Sgr~A*. Depending on the separation, there are two methods for determining the positions of the stars relative to Sgr~A*: single-beam and dual-beam astrometry. Single-beam positions are extracted from pointings in which more than one source is present in the IFOV. The distances between the stars are extracted by fitting a multi-source model to the visibility amplitude and closure phase, each of which are measured at $\approx $ ten spectral channels for six baselines. This yields  positions of the sources with respect to each other. Since Sgr~A* is visible in all our central frames, for those pointings the relative positions also are the absolute ones, that is, with respect to the mass center. If the stars are not observable in a single IFOV, we need to observe them separately and apply the dual-beam technique. For the case of two isolated stars, one interferometrically calibrates the first source with the second. The first source serves as a phase reference relative to which offsets of the second source can be measured.

\subsubsection{Single-beam astrometry}
If a star is in the same IFOV as Sgr~A* (of particular interest here in 2021 are S29 and S55), we determine the relative separation between the star and Sgr~A* by interferometric model fitting to the visibility amplitude and closure phases in the Sgr~A* pointing. This methodology is unchanged with regard to the way separations were determined in \cite{2021A&A...647A..59G}. We thus take into account the effect of phase aberrations as well as bandwidth smearing \citep{2020A&A...636L...5G}.

\subsubsection{Dual-beam astrometry}
\label{sec:A12}
If the stars are separated by more than the IFOV of GRAVITY, we measure the separation between the two sources by using one of them as the phase reference for the other target, namely, by calibrating the complex visibilities of the object of interest with the ones of the phase reference. We use S2 as the phase reference, which in its interferometric observables is consistent with a single point source. The separation between any star and S2 is determined by two vectors:
\begin{itemize}
\item the vector by which the IFOV was moved between S2 and the star. This vector is measured by the metrology system of GRAVITY monitoring the internal optical path differences 
\item the phase center offset in the S2-calibrated star observation. It is determined by fitting the visibility phase. Also for the dual-beam analysis, we account for the effect of phase aberrations and bandwidth smearing when calculating the model visibility phase.
\end{itemize}
The separation is affected by inaccuracies and systematic uncertainties of the metrology.  Such telescope-based errors are inherent to the dual-beam part of the measurement. 

Typically, we find more stars than just one in the IFOV. We thus need  to take into account the signatures induced by the additional stars in the dual-beam measurement. This occurs, for example, for the Sgr~A* pointings (where S29, S55, S62, and S300 are present), but also for the S38 pointing (with S60 and S63 being present in the IFOV). We thus fit the interstellar separations and the phase center offset simultaneously in order to take into account their degeneracies. However, the separation vectors are mostly sensitive to the visibility amplitude and the closure phase information from the visibility phase, while the phase center offsets mostly acts as an additional term in the visibility phase.
In this way we can relate all positions to our calibrator source, S2. Hence, we can relate the positions also to Sgr~A* by subtracting the star-to-S2 and S2-to-Sgr~A* separations. 

Telescope-based errors cancel out in the closure phase and therefore the relative positions of the sources are not affected by phase errors, but the visibility phases carry the information of how different IFOVs are located with respect to one another. We find that by fitting the closure phases and the visibility phases with equal weights, we minimize the effect of the telescope-based errors, while still being sensitive to the phase information. 

In order to average out the phase errors, we calibrate all $N$ frames of a given pointing with all $M$ available S2 frames individually. For each of the $N \times M$ resulting data sets, we determine the phase center position and average the resulting phase center locations. This calibration uncertainty adds a systematic uncertainty of $60~\mathrm{\mu as}$, divided by the square root of the number of available calibrations.
 
We further improve the accuracy of our phase center measurement by determining the best fit fringe-tracker and science target separation by fitting the S2 observations with a drifting point source model. This takes into account our imperfect knowledge of the separation prior to the observation. Here, we follow the concepts set out in \cite{2020A&A...636L...5G}.

\subsection{GRAVITY: Deep imaging}
\label{sec:A2}

To obtain deep, high-resolution images of the GC, we developed a new imaging code called GRAVITY-RESOLVE or $\mathrm{G^R}$ which is drawn from RESOLVE \citep{2021A&A...646A..84A}, a Bayesian imaging algorithm formulated in the framework of information field theory \citep{2019AnP...53100127E} that is custom-tailored to GRAVITY observations of the GC. Here, we  briefly outline the main ideas, (see  \cite{JULIA} for a detailed description. 

With a Bayesian forward modeling approach, we can address data sparsity and account for various instrumental effects that render the relation between image and measurement more complicated than the simple Fourier transform of the van-Cittert Zernike theorem. To this end, the algorithm formulates a prior model which permits to draw random samples, processes them with the instrumental response function, and evaluates the likelihood to compare the predicted visibilities with the actual measurement. This approach can handle the non-invertible measurement equation and allows us to work with non-linear quantities such as closure phases. The exploration of the posterior distribution is done with metric Gaussian variational inference \citep{2019arXiv190111033K}, and infers the mean image with respect to the posterior jointly with an uncertainty estimate. There are already some imaging tools available for optical/near-IR interferometry that implement a forward-modeling approach such as MIRA \citep{2008SPIE.7013E..1IT} or SQUEEZE \citep{2010SPIE.7734E..2IB}. Our code differs from those with regard to the details of the measurement equation, the prior model, and how the maximization and exploration of the posterior are performed.

In the measurement equation, we implemented all instrumental effects relevant for GRAVITY: coupling efficiency, aberration corrections \citep{2021A&A...647A..59G}, averaging over finite sized wavelength channels (also known as bandwidth smearing), and the practice in optical and near-IR interferometry to construct the complex visibility as the coherent flux over a baseline divided by the total flux of each of the two telescopes. The latter signifies that the visibility amplitude can be unity at most, but coherence loss can degrade the observed visibility from the theoretical expectation. This we account for by a self-calibration approach where we infer a time- and baseline-dependent calibration factor jointly with the image. 

An appropriate prior model is essential to address the data sparsity inherent to optical/near-IR interferometry, and we specifically tailor it to the GC observations. There, we see Sgr~A* as a point source in addition to some relatively bright stars whose approximate positions are known from orbit predictions. For those objects, we directly infer the position and brightness using a Gaussian and a log-normal prior respectively. The variability and polarization of Sgr~A* is accounted for by allowing for an independent flux value in each frame and polarization state observed. In the actual image itself, we expect to see few faint, yet unknown, point sources and thus impose the individual pixels to be independent with their brightness following an Inverse Gamma distribution. We note that all sources other than Sgr~A*, that is all non-variable sources, could in principle also be attributed to the image. However, modeling them as additional point sources improves convergence and mitigates pixelization errors.

\subsection{GNIRS: Determining radial velocities}
\label{sec:A3}
In 2021 we had four successful observations with the long-slit spectrograph GNIRS using the AO system ALTAIR at the Gemini observatory. We used the long slit in the K-Band with the $10.44\,$l/mm grating. The slit was positioned such that we observed S2 and S29 simultaneously. To calibrate the data we used the daytime calibration from the day after the observation, which contains a set of dark frames to determine a bad pixel mask, flat frames, and a wavelength calibration. Additionally, a telluric star was observed right after the observation. To determine the velocity of the stars we used template fitting with a high SNR S2 spectrum, in the same way as we extracted the SINFONI velocities (see \citealt{2018A&A...615L..15G}). We were able to detect a velocity for S2 in all four observing nights. As S29 is significantly fainter than S2 we needed excellent conditions to get a detection, which was only possible in two of the four nights.

\section{Fit details}

\begin{figure*}[t!]
\centering
%\vspace{-0.7cm}
\includegraphics[width=16cm]{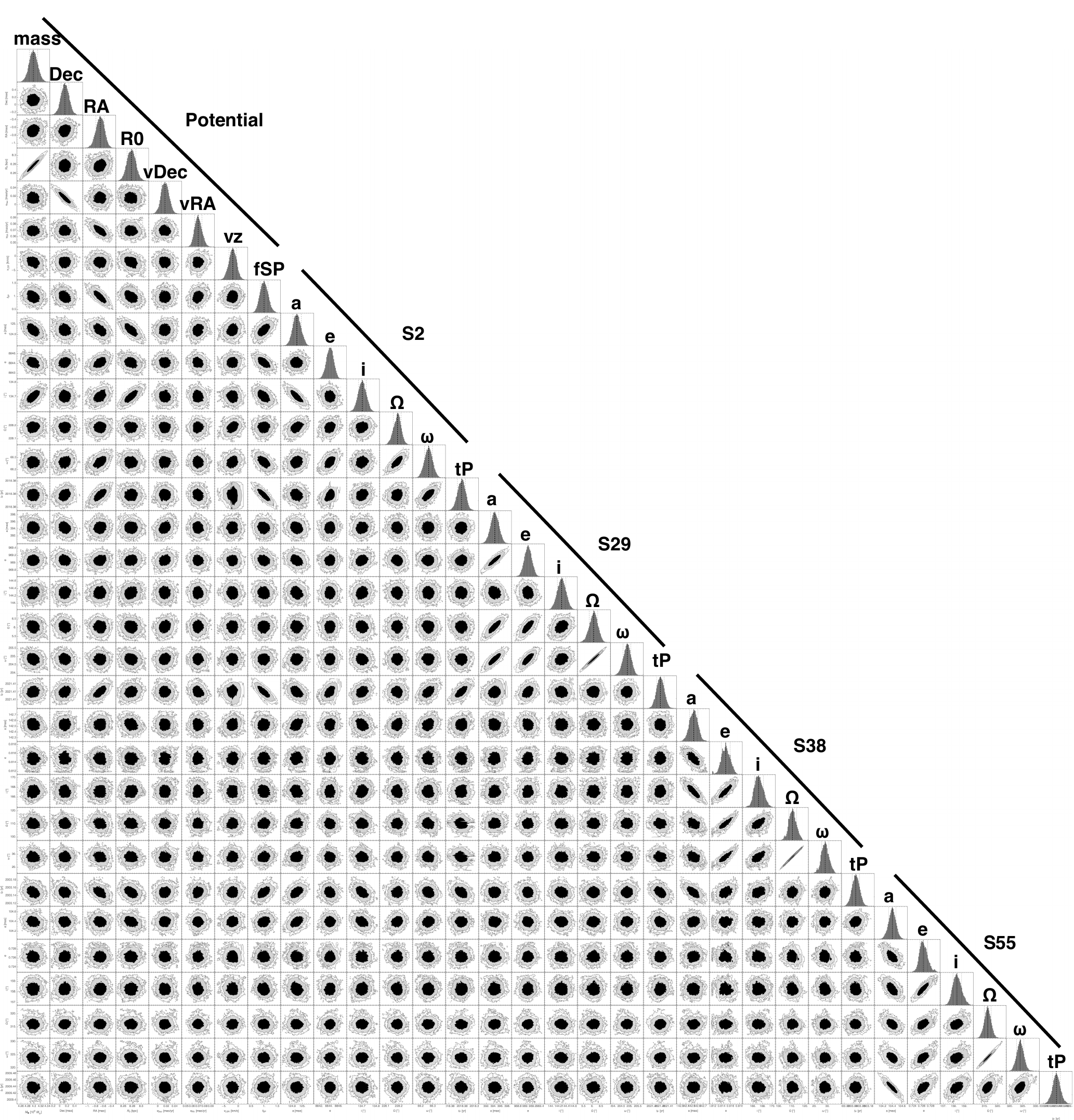}
%\vspace{-3cm}
\caption{Corner plot of the full 32-dimensional parameter space of the four-star fit, consisting of (in order) the central mass $M_\bullet$, the offsets $x_0$, $y_0$, the distance $R_0$, the velocity offsets $vx_0$, $vy_0$, $vz_0$,
the precession parameter $f_\mathrm{SP}$, followed by four times six orbital elements for each of the four stars used, in the order $a$, $e$, $i$, $\Omega$, $\omega$, $t_\mathrm{peri}$ for S2, S29, S38, S55.}
\label{figApp1}
\end{figure*}

In Table~\ref{tab1}, we give the best-fit parameters of the four-star fit-determining $f_\mathrm{SP}$, comparing also with similar fits from the literature.
Figure~\ref{figApp1} gives the full posterior of the four-star fit in the form of a corner plot.

\begin{table*}
 \caption[]{Best-fit orbit parameters of the four-star fit determining $f_\mathrm{SP}$. }
 \label{tab1}

 \begin{center}
 {\footnotesize
 \begin{tabular}{lccccccccccc}
 &\multicolumn{2}{c}{\bf this paper}&&\multicolumn{2}{c}{\bf Grav. Coll. 2020}  &&\multicolumn{2}{c}{\bf Gillessen+ 2017}  &&\multicolumn{2}{c}{\bf Do+ 2019}\\
 ~\\
 $M_\bullet \,[10^6 M_\odot]$ & 4.297 & 0.012 & & 4.261 &0.012 & & 4.280 & 0.100&&3.975 & 0.058 \\
 $R_0 \,[\mathrm{pc}]$&8277&9&&8247 &9&&8320 & 70&&7959 & 59\\
 $M_\bullet \, [10^6 M_\odot]_\mathrm{8277\,pc}$&4.297 & 0.012 && 4.292 & 0.012&&4.236 & 0.100 &&4.299 & 0.063 \\
 ~\\
 $x_0\,[\mathrm{mas}]$& -0.69 & 0.10&&-0.90&0.14&&-0.08&0.37&&1.22 & 0.32\\
 $y_0\,[\mathrm{mas}]$& 0.18 & 0.10 &&0.07 & 0.12&&-0.89&0.31&&-0.88 &0.34\\
 $vx_0\,[\mathrm{mas/yr}]$&0.066& 0.006&&0.080 & 0.010&&0.039 & 0.041&&-0.077&0.018\\
 $vy_0\,[\mathrm{mas/yr}]$&0.009&0.009&& 0.034 & 0.010&&0.058 & 0.037&&0.226 & 0.019\\
 $vz_0\,]\mathrm{km/s}]$&-1.8&1.3&& -1.6 & 1.4&& 14.2 & 3.6&&-6.2 & 3.7\\
   ~\\
   $f_\mathrm{SP}$&0.997&0.144&& 1.10 & 0.19\\
 \hline
  ~\\
 ~&\multicolumn{2}{c}{\bf S2}&&\multicolumn{2}{c}{\bf S38}\\
 ~\\
 $a\,[\mathrm{as}]$ & 0.12495 & 0.00004&&0.14254&0.00004\\
$e$ &0.88441 & 0.00006&&0.8145&0.0002\\
$i\,[^\circ]$ &134.70 & 0.03&&166.65&0.40\\
$\Omega\,[^\circ]$ &228.19 &0.03&&109.45 & 1.00\\
$\omega\,[^\circ]$ &66.25 & 0.03&&27.17 & 1.02\\
$t_\mathrm{peri}\,[\mathrm{yr}]$&2018.3789 & 0.0001 &&2022.7044 & 0.0080 \\
~\\
 ~&\multicolumn{2}{c}{\bf S29}&&\multicolumn{2}{c}{\bf S55}\\
$a\,[\mathrm{as}]$ & 0.3975 & 0.0016&&0.10440 & 0.00005\\
$e$ &0.9693 & 0.0001&&0.7267 & 0.0002\\
$i\,[^\circ]$ &144.37 & 0.07&&158.52 & 0.22\\
$\Omega\,[^\circ]$ &7.00 & 0.33&&314.94 & 1.14\\
$\omega\,[^\circ]$ &205.79 & 0.33&&322.78 & 1.13\\
$t_\mathrm{peri}\,[\mathrm{yr}]$ & 2021.4104 & 0.0002 &&2021.6940 & 0.0083\\
 \hline
~\\
\multicolumn{12}{l}{\footnotesize The line  $M_\bullet \, [10^6 M_\odot]_\mathrm{8277\,pc}$ gives the masses rescaled to a common distance of  $R_0=8277\,$pc, using $M \propto R_0^2$ \citep{2017ApJ...837...30G}.} \\
\multicolumn{12}{l}{\footnotesize For a discussion of  the differences in $R_0$ see the appendix of \cite{2021A&A...647A..59G}.
The orbital elements are meant in the sense of osculating} \\
\multicolumn{12}{l}{\footnotesize  orbit parameters, using a conversion time close to the respective apocenter times, i.e., 2010.35 for S2, 1977 for S29, 2000 for S38, and} \\
\multicolumn{12}{l}{\footnotesize  2012 for S55. The coordinate system offsets $x_0$, $y_0$ refer to the epoch 2000.0.} \\
 \end{tabular}
 }
 \end{center}
\end{table*}

\section{Additional figures}

\begin{figure*}[t!]
\centering
%\vspace{-0.7cm}
\includegraphics[width=14cm]{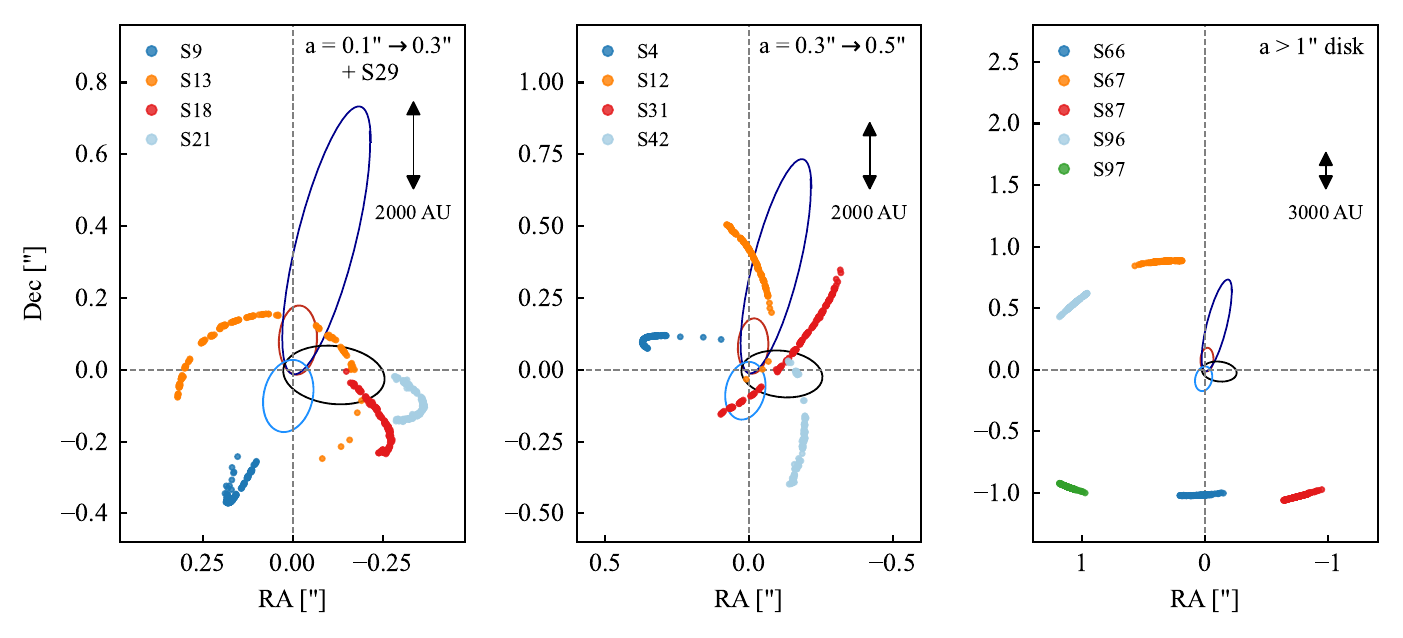}
%\vspace{-3cm}
\caption{Visualization of the orbital (astrometric) data used in determining the mass distribution in the GC. The panels group the stars according to the semi-major axes of their orbits, as indicated in the top left in each panel. For reference, we show in each panel the orbits from the four-star fit (S2: red line, S29 violet, S38 black, S55 blue),  Left: Orbital data for S9, S13, S18 and S21. Middle: S4, S12, S31 and S42. Right: S66, S67, S87, S96 and S97. These data are complemented by multi-epoch spectroscopy for the orbital fitting.}
\label{fig-orbits}
\end{figure*}

\begin{figure*}[t!]
\centering
%\vspace{-0.7cm}
\includegraphics[width=11.5cm]{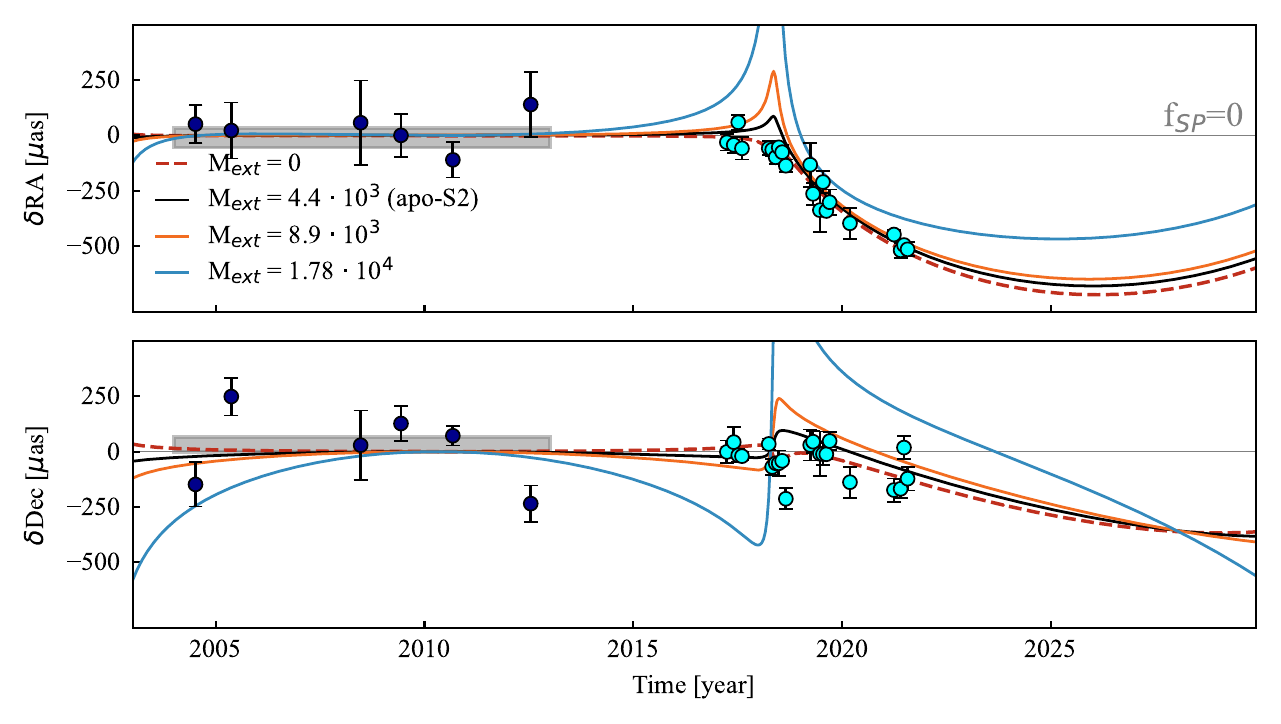}
%\vspace{-3cm}
\caption{Same data and RA-/Dec-residual plots as the left panels in Figure~\ref{fig4}, with the dashed red curve denoting the $f_\mathrm{SP}=1$ GR curve for the best fitting orbit and mass. In addition, we show orbital models with the same central mass, distance and orbital parameters but now adding an extended mass component assumed to have a Plummer shape \citep{2017ApJ...837...30G} showing the impact of adding a Plummer mass of $M_\mathrm{ext}$ within the $0.25''$ apocenter radius of S2. Black, orange and blue solid curves show the changes expected if this extended Plummer mass is 0.1, 0.3 and 0.6\% of $M_\bullet$ ($4.4\times10^3$, $8.9\times10^3$ and $1.78\times10^4 M_\odot$ within the apocenter of S2, $R_\mathrm{apo}=0.24''$).  Formal fitting shows that such an extended mass greater than about $\approx 0.1$\% of $M_\bullet$ is incompatible with the data.}
\label{fig5}
\end{figure*}

In Figure~\ref{fig-orbits} we show the orbital data of additional S-stars that were auxiliary in this work. Figure~\ref{fig5} illustrates that our S2 data are compatible at most with an extended mass component of around $0.1$\% enclosed within the S2-orbit.

\section{List of observations}
\label{app:list}
In Tables~\ref{tab:obslist}, \ref{tab:rvlist}, and \ref{tab:poslist} we list the observations used in this work. We note that most of the raw data are available from the observatories' archives, most easily found via the program ID numbers provided here. The derived astrometric and spectroscopic data set can be made available upon personal request and after discussing the terms of a collaboration on a case-by-case basis. 

\begin{table*}
 \caption[]{GRAVITY observations used in this work.}
 \label{tab:obslist}
 \begin{center}
 {\tiny
 \begin{tabular}{lcccc}
 Night&prog. ID&Instrument + Setup&type&stars\\
 \hline
2016-09-21 & 60.A-9102(A)&VLTI, GRAVITY, K-band, LOWRES & interf. pos. & S2 \\
2017-03-28 & 60.A-9102(A)& VLTI, GRAVITY, K-band, LOWRES & interf. pos. & S2 \\
2017-07-02 & 60.A-9102(A)& VLTI, GRAVITY, K-band, LOWRES & interf. pos. & S2 \\
2017-03-27 & 60.A-9102(A)& VLTI, GRAVITY, K-band, LOWRES & interf. pos. & S2 \\
2017-03-28 & 60.A-9102(A)& VLTI, GRAVITY, K-band, LOWRES & interf. pos. & S2 \\
2017-03-30 & 60.A-9102(A)& VLTI, GRAVITY, K-band, LOWRES & interf. pos. & S2 \\
2017-05-08 & 60.A-9102(A)& VLTI, GRAVITY, K-band, LOWRES & interf. pos. & S2 \\
2017-06-03 & 60.A-9102(A)& VLTI, GRAVITY, K-band, LOWRES & interf. pos. & S2 \\
2017-06-09 & 60.A-9102(A)& VLTI, GRAVITY, K-band, LOWRES & interf. pos. & S2 \\
2017-07-02 & 60.A-9102(A)& VLTI, GRAVITY, K-band, LOWRES & interf. pos. & S2 \\
2017-07-03 & 60.A-9102(A)& VLTI, GRAVITY, K-band, LOWRES & interf. pos. & S2 \\
2017-07-07 & 60.A-9102(A)& VLTI, GRAVITY, K-band, LOWRES & interf. pos. & S2 \\
2017-07-08 & 60.A-9102(A)& VLTI, GRAVITY, K-band, LOWRES & interf. pos. & S2 \\
2017-07-09 & 60.A-9102(A)& VLTI, GRAVITY, K-band, LOWRES & interf. pos. & S2 \\
2017-07-10 & 60.A-9102(A)& VLTI, GRAVITY, K-band, LOWRES & interf. pos. & S2 \\
2017-08-06 & 60.A-9102(A)& VLTI, GRAVITY, K-band, LOWRES & interf. pos. & S2 \\
2017-08-07 & 60.A-9102(A)& VLTI, GRAVITY, K-band, LOWRES & interf. pos. & S2 \\
2017-08-08 & 60.A-9102(A)& VLTI, GRAVITY, K-band, LOWRES & interf. pos. & S2 \\
2017-08-09 & 60.A-9102(A)& VLTI, GRAVITY, K-band, LOWRES & interf. pos. & S2 \\
2018-03-28 & 0100.B-0731(A)& VLTI, GRAVITY, K-band, LOWRES & interf. pos. & S2 \\
2018-03-29 & 0100.B-0731(A)& VLTI, GRAVITY, K-band, LOWRES & interf. pos. & S2 \\
2018-03-30 & 0100.B-0731(A)& VLTI, GRAVITY, K-band, LOWRES & interf. pos. & S2 \\
2018-03-31 & 0100.B-0731(A)& VLTI, GRAVITY, K-band, LOWRES & interf. pos. & S2 \\
2018-04-24 & 60.A-9102(A)& VLTI, GRAVITY, K-band, LOWRES & interf. pos. & S2 \\
2018-04-26 & 0101.B-0576(A)& VLTI, GRAVITY, K-band, LOWRES & interf. pos. & S2 \\
2018-04-28 & 0101.B-0576(A)& VLTI, GRAVITY, K-band, LOWRES & interf. pos. & S2 \\
2018-05-02 & 0101.B-0576(A)& VLTI, GRAVITY, K-band, LOWRES & interf. pos. & S2 \\
2018-05-04 & 0101.B-0576(A)& VLTI, GRAVITY, K-band, LOWRES & interf. pos. & S2 \\
2018-05-24 & 0101.B-0576(C)& VLTI, GRAVITY, K-band, LOWRES & interf. pos. & S2 \\
2018-05-27 & 0101.B-0576(C)& VLTI, GRAVITY, K-band, LOWRES & interf. pos. & S2 \\
2018-05-30 & 0101.B-0255(A)& VLTI, GRAVITY, K-band, LOWRES & interf. pos. & S2 \\
2018-06-01 & 0101.B-0576(C)& VLTI, GRAVITY, K-band, LOWRES & interf. pos. & S2 \\
2018-06-03 & 0101.B-0576(C)& VLTI, GRAVITY, K-band, LOWRES & interf. pos. & S2 \\
2018-06-23 & 0101.B-0576(D)& VLTI, GRAVITY, K-band, LOWRES & interf. pos. & S2 \\
2018-06-25 & 0101.B-0576(D)& VLTI, GRAVITY, K-band, LOWRES & interf. pos. & S2 \\
2018-06-27 & 0101.B-0576(D)& VLTI, GRAVITY, K-band, LOWRES & interf. pos. & S2 \\
2018-07-22 & 0101.B-0576(E)& VLTI, GRAVITY, K-band, LOWRES & interf. pos. & S2 \\
2018-07-25 & 0101.B-0576(E)& VLTI, GRAVITY, K-band, LOWRES & interf. pos. & S2 \\
2018-07-28 & 0101.B-0576(E)& VLTI, GRAVITY, K-band, LOWRES & interf. pos. & S2 \\
2018-08-21 & 0101.B-0576(F)& VLTI, GRAVITY, K-band, LOWRES & interf. pos. & S2 \\
2018-08-22 & 0101.B-0576(F)& VLTI, GRAVITY, K-band, LOWRES & interf. pos. & S2 \\
2018-08-27 & 0101.B-0576(F)& VLTI, GRAVITY, K-band, LOWRES & interf. pos. & S2 \\
2018-08-30 & 0101.B-0576(F)& VLTI, GRAVITY, K-band, LOWRES & interf. pos. & S2 \\
2018-08-31 & 0101.B-0576(F)& VLTI, GRAVITY, K-band, LOWRES & interf. pos. & S2 \\
2019-03-27 & 0102.B-0689(A)& VLTI, GRAVITY, K-band, LOWRES & interf. pos. & S2 \\
2019-04-21 & 0103.B-0032(A)& VLTI, GRAVITY, K-band, LOWRES & interf. pos. & S2 \\
2019-06-20 & 0103.B-0032(B)& VLTI, GRAVITY, K-band, LOWRES & interf. pos. & S2 \\
2019-07-17 & 0103.B-0032(D)& VLTI, GRAVITY, K-band, LOWRES & interf. pos. & S2 \\
2019-08-13 & 0103.B-0032(C)& VLTI, GRAVITY, K-band, LOWRES & interf. pos. & S2, S29 \\
2019-08-14 & 0103.B-0032(C)& VLTI, GRAVITY, K-band, LOWRES & interf. pos. & S2 \\
2019-08-15 & 0103.B-0032(C)& VLTI, GRAVITY, K-band, LOWRES & interf. pos. & S2 \\
2019-08-19 & 0103.B-0032(C)& VLTI, GRAVITY, K-band, LOWRES & interf. pos. & S2 \\
2019-09-12 & 0103.B-0032(D)& VLTI, GRAVITY, K-band, LOWRES & interf. pos. & S2 \\
2019-09-13 & 0103.B-0032(D)& VLTI, GRAVITY, K-band, LOWRES & interf. pos. & S2, S29 \\
2019-09-15 & 0103.B-0032(D)& VLTI, GRAVITY, K-band, LOWRES & interf. pos. & S2 \\
\hline
\multicolumn{5}{l}{\tiny   The column 'night' gives the date of the evening of the observing night.}
 \end{tabular}
 }
 \end{center}
\end{table*}

\begin{table*}
 \caption*{Table~\ref{tab:obslist} contd.}
 \begin{center}
 {\tiny
 \begin{tabular}{lcccc}
 Night&prog. ID&Instrument + Setup&type&stars\\
 \hline
2020-03-09 & 0104.B-0118(A)& VLTI, GRAVITY, K-band, LOWRES & interf. pos. & S2 \\
2021-03-27 & 106.20XY.001& VLTI, GRAVITY, K-band, LOWRES & interf. pos. & S2, S29 \\
2021-03-28 & 106.20XY.001& VLTI, GRAVITY, K-band, LOWRES & interf. pos. & S29, S55 \\
2021-03-29 & 106.20XY.001& VLTI, GRAVITY, K-band, LOWRES & interf. pos. & S29, S55 \\
2021-03-30 & 106.20XY.001& VLTI, GRAVITY, K-band, LOWRES & interf. pos. & S2, S29, S55 \\
2021-03-31 & 1103.B-0626(E)& VLTI, GRAVITY, K-band, LOWRES & interf. pos. & S2, S29, S38, S55 \\
2021-05-22 & 105.20B2.002& VLTI, GRAVITY, K-band, LOWRES & interf. pos. & S2, S29, S38, S55 \\
2021-05-23 & 105.20B2.002& VLTI, GRAVITY, K-band, LOWRES & interf. pos. & S2, S29, S38, S55 \\
2021-05-24 & 105.20B2.002& VLTI, GRAVITY, K-band, LOWRES & interf. pos. & S2, S29, S55 \\
2021-05-25 & 105.20B2.002& VLTI, GRAVITY, K-band, LOWRES & interf. pos. & S29, S55 \\
2021-05-28 & 60.A-9256(A)& VLTI, GRAVITY, K-band, LOWRES & interf. pos. & S2, S29, S55 \\
2021-05-29 & 105.20B2.002& VLTI, GRAVITY, K-band, LOWRES & interf. pos. & S2, S29, S55 \\
2021-05-30 & 105.20B2.002& VLTI, GRAVITY, K-band, LOWRES & interf. pos. & S2, S29, S38, S55 \\
2021-06-19 & 105.20B2.003& VLTI, GRAVITY, K-band, LOWRES & interf. pos. & S29, S55 \\
2021-06-24 & 105.20B2.003& VLTI, GRAVITY, K-band, LOWRES & interf. pos. & S2, S29, S55 \\
2021-06-25 & 105.20B2.003& VLTI, GRAVITY, K-band, LOWRES & interf. pos. & S2, S29, S55 \\
2021-07-25 & 1103.B-0626(C)& VLTI, GRAVITY, K-band, LOWRES & interf. pos. & S2, S29, S38, S55 \\
2021-07-26 & 1103.B-0626(C)& VLTI, GRAVITY, K-band, LOWRES & interf. pos. & S2, S29, S38, S55 \\
2021-07-27 & 105.20B2.004& VLTI, GRAVITY, K-band, LOWRES & interf. pos. & S2, S29, S55 \\
2021-07-29 & 105.20B2.004& VLTI, GRAVITY, K-band, LOWRES & interf. pos. & S2, S29, S55 
 \end{tabular}
 }
 \end{center}
\end{table*}

\begin{table*}
 \caption[]{Radial velocity observations used in this work. }
 \label{tab:rvlist}
 \begin{center}
 {\tiny
 \begin{tabular}{lcccc}
 Night&prog. ID / publication &Instrument + Setup&type&stars\\
 \hline
2000-06-22 & \cite{2019Sci...365..664D} & Keck, NIRSPEC & spectr. rad. vel. & S2\\
2002-06-01 & \cite{2019Sci...365..664D} & Keck, NIRC2 & spectr. rad. vel. & S2\\
2002-06-02 & \cite{2019Sci...365..664D} & Keck, NIRC2 & spectr. rad. vel. & S2\\
2003-05-09 & \cite{2003ApJ...597L.121E} & VLT, NACO & AO spectr. rad. vel. & S2\\
2003-06-12 &  \cite{2003ApJ...597L.121E} & VLT, NACO & AO spectr. rad. vel. & S2\\
2003-04-07 & 70.A-0229(A) & VLT, SPIFFI, K-band & spectr. rad. vel. & S2\\
2004-07-14 & 60.A-9235(A) & VLT, SINFONI, H+K-band & AO spectr. rad. vel. & S2\\
2004-07-16 &  60.A-9235(A)  & VLT, SINFONI, K-band & AO spectr. rad. vel. & S2\\
2004-08-18 &  073.B-9017(A) & VLT, SINFONI, K-band & AO spectr. rad. vel. & S2\\
2005-02-27 & 60.A-9235(A) & VLT, SINFONI, K-band & AO spectr. rad. vel. & S2\\
2005-03-19 & 074.B-9014(A)  & VLT, SINFONI, K-band & AO spectr. rad. vel. & S2\\
2005-06-17 & 075.B-0547(B) & VLT, SINFONI, K-band & AO spectr. rad. vel. & S2\\
2005-10-06 & 076.B-0259(A) & VLT, SINFONI, H+K-band & AO spectr. rad. vel. & S2\\
2006-03-16 & 076.B-0259(B)  & VLT, SINFONI, H+K-band & AO spectr. rad. vel. & S2\\
2006-04-22 & 077.B-0503(B) & VLT, SINFONI, H+K-band & AO spectr. rad. vel. & S2\\
2006-08-17 &  077.B-0503(C) & VLT, SINFONI, H+K-band & AO spectr. rad. vel. & S2\\
2007-03-26 &  078.B-0520(A) & VLT, SINFONI, H+K-band & AO spectr. rad. vel. & S2\\
2007-04-21 & 179.B-0261(F)  & VLT, SINFONI, H+K-band & AO spectr. rad. vel. & S2\\
2007-07-23 & 179.B-0261(Z)  & VLT, SINFONI, H+K-band & AO spectr. rad. vel. & S2, S29\\
2007-09-03 &  179.B-0261(K) & VLT, SINFONI, H+K-band & AO spectr. rad. vel. & S2\\
2008-04-07 &  081.B-0568(A) & VLT, SINFONI, H+K-band & AO spectr. rad. vel. & S2, S29, S38\\
2009-05-24 & 183.B-0100(B)  & VLT, SINFONI, H+K-band & AO spectr. rad. vel. & S2\\
2009-08-20 & 183.B-0100(E) & VLT, SINFONI, H+K-band & AO spectr. rad. vel. & S2\\
2010-05-11 & 183.B-0100(O)  & VLT, SINFONI, H+K-band & AO spectr. rad. vel. & S2, S29\\
2010-06-09 & 183.B-1004(B)       & VLT, SINFONI, H+K-band & AO spectr. rad. vel. & S2\\
2010-07-09 & 179.B-0932(C) & VLT, SINFONI, H+K-band & AO spectr. rad. vel. & S2\\
2011-04-27 &087.B-0117(I)  & VLT, SINFONI, H+K-band & AO spectr. rad. vel. & S2\\
2011-05-02 & 087.B-0117(I) & VLT, SINFONI, H+K-band & AO spectr. rad. vel. & S2\\
2011-05-14 & 087.B-0117(I) & VLT, SINFONI, H+K-band & AO spectr. rad. vel. & S2\\
2011-07-27 & 087.B-0117(J)  & VLT, SINFONI, H+K-band & AO spectr. rad. vel. & S2\\
2012-03-19 & 288.B-5040(A) & VLT, SINFONI, H+K-band & AO spectr. rad. vel. & S2\\
2012-05-04 & 087.B-0117(J) & VLT, SINFONI, H+K-band & AO spectr. rad. vel. & S2\\
2012-06-29 & 288.B-5040(A  & VLT, SINFONI, H+K-band & AO spectr. rad. vel. & S2, S29, S38\\
2012-09-13 & 087.B-0117(J) & VLT, SINFONI, H+K-band & AO spectr. rad. vel. & S2\\
2013-04-04 & 091.B-0088(A) & VLT, SINFONI, H+K-band & AO spectr. rad. vel. & S2, S29\\
2013-04-13 & 091.B-0081(G) & VLT, SINFONI, H+K-band & AO spectr. rad. vel. & S2\\
2013-05-11&\cite{2016ApJ...830...17B} &Keck, NIRC2   & AO spectr. rad. vel. & S38\\
2013-08-27 & 091.B-0088(B)  & VLT, SINFONI, H+K-band & AO spectr. rad. vel. & S2\\
2013-09-21 & 60.A-9800(K) & VLT, SINFONI, H+K-band & AO spectr. rad. vel. & S2\\
2014-03-05 & 092.B-0398(B) & VLT, SINFONI, H+K-band & AO spectr. rad. vel. & S2, S55\\
2014-04-05 & 093.B-0218(A) & VLT, SINFONI, H+K-band & AO spectr. rad. vel. & S2\\
2014-04-23 &  093.B-0218(A)& VLT, SINFONI, H+K-band & AO spectr. rad. vel. & S2, S29\\
2014-05-07 & 093.B-0217(F) & VLT, SINFONI, H+K-band & AO spectr. rad. vel. & S2\\
2014-05-26 &  092.B-0398(D)& VLT, SINFONI, H+K-band & AO spectr. rad. vel. & S2\\
2014-06-04 &  092.B-0398(D)& VLT, SINFONI, H+K-band & AO spectr. rad. vel. & S2\\
2014-06-09 & 092.B-0398(A) & VLT, SINFONI, H+K-band & AO spectr. rad. vel. & S2\\
2014-07-07 & 092.B-0398(D) & VLT, SINFONI, H+K-band & AO spectr. rad. vel. & S2, S29, S38, S55\\
2014-08-17 & 093.B-0218(D)  & VLT, SINFONI, H+K-band & AO spectr. rad. vel. & S2\\
2014-08-29 &093.B-0218(B)  & VLT, SINFONI, H+K-band & AO spectr. rad. vel. & S2\\
2015-04-20 & 093.B-0218(D) & VLT, SINFONI, H+K-band & AO spectr. rad. vel. & S2, S38\\
2015-04-27 & 594.B-0498(Q) & VLT, SINFONI, H+K-band & AO spectr. rad. vel. & S2\\
2015-05-19 & 093.B-0218(D) & VLT, SINFONI, H+K-band & AO spectr. rad. vel. & S2\\
2015-07-07 & 095.B-0640(B) & VLT, SINFONI, H+K-band & AO spectr. rad. vel. & S2\\
2015-09-15 & 093.B-0218(D & VLT, SINFONI, H+K-band & AO spectr. rad. vel. & S2\\
2016-04-13 & 594.B-0498(R) & VLT, SINFONI, H+K-band & AO spectr. rad. vel. & S2, S29, S38\\
2016-07-10 & 097.B-0050(A) & VLT, SINFONI, H+K-band & AO spectr. rad. vel. & S2\\
2017-03-18 & 598.B-0043(A) & VLT, SINFONI, H+K-band & AO spectr. rad. vel. & S2\\
2017-04-05 & 598.B-0043(B) & VLT, SINFONI, H+K-band & AO spectr. rad. vel. & S2, S29\\
2017-04-16 & 598.B-0043(B) & VLT, SINFONI, H+K-band & AO spectr. rad. vel. & S2\\
2017-05-19 &  299.B-5014(A) & VLT, SINFONI, H+K-band & AO spectr. rad. vel. & S2, S29\\
2017-06-01 & 299.B-5014(A)  & VLT, SINFONI, H+K-band & AO spectr. rad. vel. & S2\\
2017-06-28 & 299.B-5014(A)  & VLT, SINFONI, H+K-band & AO spectr. rad. vel. & S2\\
2017-07-19 & 299.B-5014(A) & VLT, SINFONI, H+K-band & AO spectr. rad. vel. & S2, S29\\
2017-07-28 & 598.B-0043(C) & VLT, SINFONI, H+K-band & AO spectr. rad. vel. & S2\\
2017-08-18 & 299.B-5014(A) & VLT, SINFONI, H+K-band & AO spectr. rad. vel. & S2\\
2017-09-14 & 299.B-5014(A) & VLT, SINFONI, H+K-band & AO spectr. rad. vel. & S2\\
2017-09-27 & 099.B-0275(B)  & VLT, SINFONI, H+K-band & AO spectr. rad. vel. & S2\\
2017-10-18 & 299.B-5056(A)  & VLT, SINFONI, H+K-band & AO spectr. rad. vel. & S2\\
\hline
\multicolumn{5}{l}{\tiny   The column 'night' gives the date of the evening of the observing night. In some cases, data from different nights,} \\
\multicolumn{5}{l}{\tiny   not too much separated in time, have been co-added for signal-to-noise reasons.}
 \end{tabular}
 }
 \end{center}
\end{table*}

\begin{table*}
 \caption*{Table~\ref{tab:rvlist} contd.}
 \begin{center}
 {\tiny
 \begin{tabular}{lcccc}
 Night&prog. ID / publication &Instrument + Setup&type&stars\\
 \hline
2018-02-15 & 299.B-5056(B)  & VLT, SINFONI, H+K-band & AO spectr. rad. vel. & S2\\
2018-03-23 & 598.B-0043(D)  & VLT, SINFONI, H+K-band & AO spectr. rad. vel. & S2, S29\\
2018-03-25 & 598.B-0043(D)  & VLT, SINFONI, H+K-band & AO spectr. rad. vel. & S2\\
2018-04-08 & 0101.B-0195(B) & VLT, SINFONI, H+K-band & AO spectr. rad. vel. & S2\\
2018-04-27 & 598.B-0043(E)  & VLT, SINFONI, H+K-band & AO spectr. rad. vel. & S2, S29, S38\\
2018-04-29 & 598.B-0043(E)  & VLT, SINFONI, H+K-band & AO spectr. rad. vel. & S2\\
2018-05-03 & 598.B-0043(E)  & VLT, SINFONI, H+K-band & AO spectr. rad. vel. & S2\\
2018-05-16 &  0101.B-0195(C) & VLT, SINFONI, H+K-band & AO spectr. rad. vel. & S2\\
2018-05-19 & 0101.B-0195(D) & VLT, SINFONI, H+K-band & AO spectr. rad. vel. & S2\\
2018-05-27 & 0101.B-0195(E) & VLT, SINFONI, H+K-band & AO spectr. rad. vel. & S2\\
2018-05-29 & 598.B-0043(F) & VLT, SINFONI, H+K-band & AO spectr. rad. vel. & S2, S29, S38\\
2018-06-02 &598.B-0043(F)  & VLT, SINFONI, H+K-band & AO spectr. rad. vel. & S2\\
2018-06-06 & 598.B-0043(F) & VLT, SINFONI, H+K-band & AO spectr. rad. vel. & S2\\
2018-06-22 & 0101.B-0195(F) & VLT, SINFONI, H+K-band & AO spectr. rad. vel. & S2, S29\\
2018-06-24 &  598.B-0043(G)& VLT, SINFONI, H+K-band & AO spectr. rad. vel. & S2\\
2018-07-02 & 598.B-0043(G) & VLT, SINFONI, H+K-band & AO spectr. rad. vel. & S2\\
2018-07-08 & 0101.B-0195(G)       & VLT, SINFONI, H+K-band & AO spectr. rad. vel. & S2\\
2018-07-27 &598.B-0043(H)  & VLT, SINFONI, H+K-band & AO spectr. rad. vel. & S2\\
2018-08-02 &  598.B-0043(H)& VLT, SINFONI, H+K-band & AO spectr. rad. vel. & S2\\
2018-08-05 & 598.B-0043(H) & VLT, SINFONI, H+K-band & AO spectr. rad. vel. & S2\\
2018-08-18 &598.B-0043(I)   & VLT, SINFONI, H+K-band & AO spectr. rad. vel. & S2\\
2018-08-19 &  598.B-0043(J) & VLT, SINFONI, H+K-band & AO spectr. rad. vel. & S2\\
2018-09-27 & 598.B-0043(J)  & VLT, SINFONI, H+K-band & AO spectr. rad. vel. & S2\\
2019-04-19 & 0103.B-0026(B) & VLT, SINFONI, H+K-band & AO spectr. rad. vel. & S2\\
2019-04-29 & 0103.B-0026(B) & VLT, SINFONI, H+K-band & AO spectr. rad. vel. & S2\\
2019-05-22 & 5102.B-0086(Q)  & VLT, SINFONI, H+K-band & AO spectr. rad. vel. & S2, S29, S38\\
2019-06-01 & 0103.B-0026(F) & VLT, SINFONI, H+K-band & AO spectr. rad. vel. & S2\\
2019-06-07 & 5102.B-0086(Q) & VLT, SINFONI, H+K-band & AO spectr. rad. vel. & S2\\
2019-06-13 & 0103.B-0026(D) & VLT, SINFONI, H+K-band & AO spectr. rad. vel. & S2\\
2021-04-28 & GN-2021A-DD-104 & Gemini, GNIRS & AO spectr. rad. vel. & S2, S29\\
2021-05-18 & GN-2021A-DD-104 & Gemini, GNIRS & AO spectr. rad. vel. & S2, S29\\
2021-06-12 &GN-2021A-DD-104  & Gemini, GNIRS & AO spectr. rad. vel. & S2\\
2021-08-02 &  GN-2021A-FT-115 & Gemini, GNIRS & AO spectr. rad. vel. & S2
 \end{tabular}
 }
 \end{center}
\end{table*}

\begin{table*}
 \caption[]{Astrometric observations used in this work.}
 \label{tab:poslist}
 \begin{center}
 {\tiny
 \begin{tabular}{lcccc}
 Night&prog. ID / publication &Instrument + Setup&type&stars\\
 \hline
1992-03 & NA  & NTT, SHARP, K-band & AO pos. & S2\\
1994-04 & NA & NTT, SHARP, K-band & AO pos. & S2\\
1995-07 & NA & NTT, SHARP, K-band & AO pos. & S2\\
1996-04 &NA  & NTT, SHARP, K-band & AO pos. & S2\\
1996-06 &  NA& NTT, SHARP, K-band & AO pos. & S2\\
1997-07 & NA & NTT, SHARP, K-band & AO pos. & S2\\
1999-06 & NA& NTT, SHARP, K-band & AO pos. & S2\\
2000-06 & NA & NTT, SHARP, K-band & AO pos. & S2\\
2000-07 & NA & Gemini, K-band & AO pos. & S2\\
2001-07 & NA & NTT, SHARP, K-band & AO pos. & S2\\
2002-05-03& 60.A-9026(A) & VLT, NACO, K-band & AO pos. & S29 \\
2002-05-30& 60.A-9026(A) & VLT, NACO, K-band & AO pos. & S29 \\
2002-05-31&60.A-9026(A)  & VLT, NACO, K-band & AO pos. & S29 \\
2002-06-01& 60.A-9026(A) & VLT, NACO, K-band & AO pos. & S29 \\
2002-07-31 & 60.A-9026(A)  & VLT, NACO, K-band & AO pos. & S2, S29\\
2002-08-29 & 60.A-9026(A)  & VLT, NACO, H-, K-band & AO pos. & S2, S29\\
2003-03-20 & 70.B-0649(B)  &VLT, NACO, H-band & AO pos. & S29\\ 
2003-05-09 & 71.B-0077(A)   &VLT, NACO, H-band & AO pos. & S29\\ 
2003-05-11 & 71.B-0077(A)   &VLT, NACO, K-band & AO pos. & S29\\ 
2003-06-13 & 71.B-0078(A)   &VLT, NACO, K-band & AO pos. & S29\\ 
2003-06-15 & 71.B-0078(A)   &VLT, NACO, H-, K-band & AO pos. & S29\\ 
2003-05-16 & 71.B-0078(A)   &VLT, NACO, H-, K-band & AO pos. & S29\\ 
2003-07-21 & 71.B-0077(C) & VLT, NACO, H-band & AO pos. & S2\\
2003-09-05 & 71.B-0077(D) & VLT, NACO, H-band & AO pos. & S2, S29\\
2003-09-06 & 71.B-0077(D) & VLT, NACO, H-band & AO pos. & S2\\
2003-10-06 & 072.B-0285(A) & VLT, NACO, K-band & AO pos. & S2, S29\\
2004-03-28 & 072.B-0285(B) & VLT, NACO, H-band & AO pos. & S2, 29, S38\\
2004-04-28 & 073.B-0085(A) & VLT, NACO, H-band & AO pos. & S2, S29, S38\\
2004-05-06 & 073.B-0084(B)  & VLT, NACO, K-band & AO pos. & S2, S29\\
2004-06-10 & 073.B-0084(A)  & VLT, NACO, H-band & AO pos. & S2, S29\\
2004-07-05 & 073.B-0775(A) & VLT, NACO, H-, K-band & AO pos. & S2, S29, S38, S55\\
2004-07-07 & 073.B-0775(A) & VLT, NACO, H-, K-band & AO pos. & S2, S29, S55\\
2004-07-28 & 273.B-5023(C) & VLT, NACO, H-, K-band & AO pos. & S2, S29\\
2004-08-30 & 073.B-0775(B) & VLT, NACO, K-band & AO pos. & S2, S29\\
2004-09-01 & 073.B-0775(B)  & VLT, NACO, K-band & AO pos. & S2, S29, S55\\
2004-09-23 & 073.B-0085(C) & VLT, NACO, H-, K-band & AO pos. & S2, S29, S55\\
2005-04-08 & 073.B-0085(I) & VLT, NACO, K-band & AO pos. & S2, S29, S38\\
2005-05-13 & 073.B-0085(D) & VLT, NACO, K-band & AO pos. & S2, S29, S38\\
2005-05-15 & 073.B-0085(D) & VLT, NACO, K-band & AO pos. & S2, S29, S38\\
2005-05-16 & 073.B-0085(D) & VLT, NACO, K-band & AO pos. & S2, S29, S38\\
2005-06-19 & 073.B-0085(F) & VLT, NACO, K-band & AO pos. & S2\\
2005-07-27 & 075.B-0093(C) & VLT, NACO, K-band & AO pos. & S2\\
2005-07-29 & 075.B-0093(B) & VLT, NACO, K-band & AO pos. & S2, S29\\
2005-09-03 & 073.B-0085(H) & VLT, NACO, K-band & AO pos. & S38\\
2006-04-29 & 077.B-0014(A) & VLT, NACO, H-band & AO pos. & S29, S38\\
2006-05-30 & 077.B-0014(B) & VLT, NACO, H-band & AO pos. & S29, S38\\
2006-06-29 & 077.B-0014(C) & VLT, NACO, H-band & AO pos. & S29\\
2006-08-02 & 077.B-0014(D) & VLT, NACO, H-band & AO pos. & S29, S38\\
2006-08-28 & 077.B-0014(E) & VLT, NACO, H-band & AO pos. & S29\\
2006-09-23 & 077.B-0014(F) & VLT, NACO, K-band & AO pos. & S29, S38\\
2006-09-24 & 077.B-0014(F) & VLT, NACO, K-band & AO pos. & S29\\
2006-10-03 & 078.B-0136(A) & VLT, NACO, K-band & AO pos. & S29, S38\\
2006-10-13 & 078.B-0136(A) & VLT, NACO, K-band & AO pos. & S29, S38\\
2006-10-20 & 078.B-0136(A) & VLT, NACO, K-band & AO pos. & S38\\
2007-03-04 & 078.B-0136(B) & VLT, NACO, K-band & AO pos. & S29, S38\\
2007-03-16 & 078.B-0136(B) & VLT, NACO, H-, K-band & AO pos. & S29, S38, S55\\
2007-03-19 & 078.B-0136(B) & VLT, NACO, K-band & AO pos. & S29, S38, S55\\
2007-04-02 & 179.B-0261(A) & VLT, NACO, K-band & AO pos. & S29, S38\\
2007-04-03 & 179.B-0261(A) & VLT, NACO, H-, K-band & AO pos. & S29, S38, S55\\
2007-05-19 & 179.B-0261(H) & VLT, NACO, K-band & AO pos. & S29, S38\\
2007-06-16 & 179.B-0261(H) & VLT, NACO, K-band & AO pos. & S29, S38\\
2007-06-18 & 179.B-0261(H) & VLT, NACO, K-band & AO pos. & S38\\
2007-07-18 & 179.B-0261(D) & VLT, NACO, H-band & AO pos. & S38\\
2007-07-20 & 179.B-0261(D) & VLT, NACO, H-band & AO pos. & S29, S38, S55\\
2007-09-08 & 179.B-0261(J) & VLT, NACO, H-band & AO pos. & S29, S38, S55\\
2007-09-09 & 179.B-0261(J) & VLT, NACO, K-band & AO pos. & S29, S38\\
2007-09-10 & 179.B-0261(J) & VLT, NACO, K-band & AO pos. & S29, S38 \\
\hline
\multicolumn{5}{l}{\tiny   The column 'night' gives the date of the evening of the observing night.}
 \end{tabular}
 }
 \end{center}
\end{table*}

\begin{table*}
 \caption*{Table~\ref{tab:poslist} contd.}
  \begin{center}
 {\tiny
 \begin{tabular}{lcccc}
 Night&prog. ID / publication &Instrument + Setup&type&stars\\
 \hline
2008-02-23 & 179.B-0261(L) & VLT, NACO, K-band & AO pos. & S2\\
2008-03-13 & 179.B-0261(L) & VLT, NACO, K-band & AO pos. & S2\\
2008-04-07 & 179.B-0261(M)  & VLT, NACO, K-band & AO pos. & S2, S38\\
2008-06-16 & 179.B-0261(T) & VLT, NACO, K-band & AO pos. & S2\\
2008-06-21 & 179.B-0261(T) & VLT, NACO, H-band & AO pos. & S2, S29, S38\\
2008-06-22 & 179.B-0261(T) & VLT, NACO, H-band & AO pos. & S2, S29\\
2008-08-05 & 179.B-0261(N) & VLT, NACO, K-band & AO pos. & S2, S29, S38\\
2008-08-07 & 179.B-0261(N) & VLT, NACO, K-band & AO pos. & S2, S38\\
2008-09-16 & 179.B-0261(U) & VLT, NACO, K-band & AO pos. & S2, S29, S38\\
2009-03-09 & 179.B-0261(X)& VLT, NACO, K-band & AO pos. & S2, S38\\
2009-04-09 & 179.B-0261(W) & VLT, NACO, K-band & AO pos. & S2, S29, S38\\
2009-04-19 & 179.B-0261(W) & VLT, NACO, K-band & AO pos. & S2, S29, S38\\
2009-04-20 & 179.B-0261(W) & VLT, NACO, K-band & AO pos. & S2, S29, S38\\
2009-05-02 & 183.B-0100(G) & VLT, NACO, H-, K-band & AO pos. & S2, S29, S38\\
2009-05-15 & 183.B-0100(G) & VLT, NACO, K-band & AO pos. & S2, S29, S38\\
2009-07-02 & 183.B-0100(D) & VLT, NACO, H-band & AO pos. & S2, S29, S38\\
2009-07-03 & 183.B-0100(D) & VLT, NACO, K-band & AO pos. & S2, S29\\
2009-07-22 & 183.B-0100(H) & VLT, NACO, H-, K-band & AO pos. & S2, S29, S38\\
2009-08-09 & 183.B-0100(I)  & VLT, NACO, K-band & AO pos. & S2, S29, S38\\
2009-09-19 & 183.B-0100(J) & VLT, NACO, K-band & AO pos. & S2, S29, S38\\
2009-10-10 &  183.B-0100(K)& VLT, NACO, K-band & AO pos. & S2\\
2010-03-27 & 183.B-0100(L) & VLT, NACO, K-band & AO pos. & S2, S38, S55\\
2010-03-29 & 183.B-0100(L) & VLT, NACO, H-, K-band & AO pos. & S2, S38, S55\\
2010-03-31 & 183.B-0100(L) & VLT, NACO, K-band & AO pos. & S2, S38, S55\\
2010-05-09 & 183.B-0100(T) & VLT, NACO, K-band & AO pos. & S2, S38, S55\\
2010-06-12 & 183.B-0100(T) & VLT, NACO, K-band & AO pos. & S2, S38, S55\\
2010-06-16 & 183.B-0100(S)  & VLT, NACO, H-, K-band & AO pos. & S2, S38, S55\\
2010-06-18 & 183.B-0100(U) & VLT, NACO, K-band & AO pos. & S2, S38, S55\\
2010-08-13 & 183.B-0100(M) & VLT, NACO, H-band & AO pos. & S2, S38, S55\\
2010-08-14 &  183.B-0100(M) & VLT, NACO, K-band & AO pos. & S2, S38, S55\\
2010-08-15 & 183.B-0100(M)  & VLT, NACO, K-band & AO pos. & S2, S38, S55\\
2010-08-16 & 183.B-0100(M)  & VLT, NACO, K-band & AO pos. & S2, S38, S55\\
2010-09-04 & 183.B-0100(V) & VLT, NACO, K-band & AO pos. & S2, S38\\
2010-09-05 &183.B-0100(V)  & VLT, NACO, K-band & AO pos. & S2, S38, S55 \\
2011-03-28 & 183.B-0100(X) & VLT, NACO, K-band & AO pos. & S2, S38\\
2011-03-29 & 087.B-0280(D) & VLT, NACO, K-band & AO pos. & S2, S38\\
2011-03-30 &  183.B-0100(X)& VLT, NACO, K-band & AO pos. & S2, S38\\
2011-04-01 &  183.B-0100(X)& VLT, NACO, K-band & AO pos. & S2, S38\\
2011-04-24 & 183.B-0100(XW) & VLT, NACO, H-, K-band & AO pos. & S2, S38\\
2011-04-25 & 183.B-0100(W) & VLT, NACO, K-band & AO pos. & S2, S38\\
2011-05-03 & 087.B-0117(B) & VLT, NACO, K-band & AO pos. & S2, S38\\
2011-06-11 & 087.B-0117(D) & VLT, NACO, K-band & AO pos. & S2, S38\\
2011-07-21 & 087.B-0117(G) & VLT, NACO, K-band & AO pos. & S2\\
2011-08-12 & 087.B-0117(G) & VLT, NACO, K-band & AO pos. & S2, S38\\
2011-09-09 & 087.B-0117(H) & VLT, NACO, K-band & AO pos. & S2, S38\\
2011-09-11 & 087.B-0117(H) & VLT, NACO, H-, K-band & AO pos. & S2, S38, S55\\
2011-09-12 & 087.B-0117(H) & VLT, NACO, K-band & AO pos. & S2, S38, S55\\
2011-09-21 & 087.B-0117(H) & VLT, NACO, K-band & AO pos. & S2, S38\\
2012-03-14 & 088.B-1038(A) & VLT, NACO, K-band & AO pos. & S2, S38\\
2012-05-03 & 088.B-0308(A)  & VLT, NACO, K-band & AO pos. & S2, S38, S55\\
2012-06-30 & 089.B-0162(B) & VLT, NACO, K-band & AO pos. & S2, S38, S55\\
2012-07-13 & 089.B-0162(C) & VLT, NACO, K-band & AO pos. & S2, S38, S55\\
2012-07-17 & 089.B-0162(A) & VLT, NACO, K-band & AO pos. & S2, S38, S55\\
2012-07-20 &  088.B-0308(B) & VLT, NACO, H-, K-band & AO pos. & S2, S38, S55\\
2012-08-08 & 089.B-0162(D) & VLT, NACO, K-band & AO pos. & S2, S38, S55\\
2012-09-12 &089.B-0162(E)  & VLT, NACO, K-band & AO pos. & S2, S38, S55\\
2013-03-28 &089.B-0162(E)& VLT, NACO, K-band & AO pos. & S38, S55\\
2013-04-25& 091.B-0081(C)& VLT, NACO, K-band & AO pos. & S38, S55\\
2013-05-13& 091.B-0081(E)& VLT, NACO, K-band & AO pos. & S38, S55\\
2013-06-02& 091.B-0081(B)& VLT, NACO, H-band & AO pos. & S38, S55\\
2013-06-08&091.B-0081(A)& VLT, NACO, H-band & AO pos. & S38\\
2013-06-29&091.B-0081(D)& VLT, NACO, H-band & AO pos. & S38, S55\\
2013-07-02&091.B-0081(D)& VLT, NACO, H-band & AO pos. & S38, S55\\
2013-08-02&091.B-0081(F)& VLT, NACO, H-band & AO pos. & S38, S55\\
2013-08-03&091.B-0081(F)& VLT, NACO, H-band & AO pos. & S38\\
2013-08-13&091.B-0081(F)& VLT, NACO, H-band & AO pos. & S38, S55\\
2015-06-07& 093.B-0217(B) & VLT, NACO, K-band & AO pos. & S38\\
2015-07-07& 095.B-0301(A)& VLT, NACO, K-band & AO pos. & S38
 \end{tabular}
 }
 \end{center}
\end{table*}

\begin{table*}
 \caption*{Table~\ref{tab:poslist} contd.}
  \begin{center}
 {\tiny
 \begin{tabular}{lcccc}
 Night&prog. ID / publication &Instrument + Setup&type&stars\\
 \hline
2016-03-21 & 594.B-0498(J) & VLT, NACO, K-band & AO pos. & S2\\
2016-04-14 & 594.B-0498(K) & VLT, NACO, K-band & AO pos. & S2, S38\\
2016-04-28 & 594.B-0498(K) & VLT, NACO, K-band & AO pos. & S2\\
2016-05-14 & 594.B-0498(K) & VLT, NACO, K-band & AO pos. & S2\\
2016-07-10 & 594.B-0498(K) & VLT, NACO, K-band & AO pos. & S2\\
2016-07-12 & 594.B-0498(K) & VLT, NACO, K-band & AO pos. & S2\\
2017-03-25 & 598.B-0043(K) & VLT, NACO, K-band & AO pos. & S2\\
2017-04-27 & 598.B-0043(M) & VLT, NACO, K-band & AO pos. & S2\\
2017-04-30 & 598.B-0043(L) & VLT, NACO, K-band & AO pos. & S2\\
2017-06-01 & 598.B-0043(L) & VLT, NACO, H-, K-band & AO pos. & S2, S38\\
2017-06-15 & 598.B-0043(L) & VLT, NACO, K-band & AO pos. & S2\\
2017-07-26 & 598.B-0043(LM)  & VLT, NACO, H-,K-band & AO pos. & S2\\
2017-08-16 & 598.B-0043(L) & VLT, NACO, K-band & AO pos. & S2\\
2017-08-21 &598.B-0043(L)  & VLT, NACO, K-band & AO pos. & S2, S38\\
2018-03-21 & 598.B-0043(N) & VLT, NACO, K-band & AO pos. & S29\\
2018-04-29 & 598.B-0043(O)& VLT, NACO, K-band & AO pos. & S29\\
2018-05-11 & 0101.B-0195(A) & VLT, NACO, K-band & AO pos. & S29\\
2018-05-15 & 0101.B-0195(A) & VLT, NACO, K-band & AO pos. & S29\\       
2018-05-24 & 598.B-0043(O) & VLT, NACO, K-band & AO pos. & S29\\        
2018-06-08& 598.B-0043(O)& VLT, NACO, K-band & AO pos. & S29\\  
2018-06-12& 598.B-0043(O)& VLT, NACO, K-band & AO pos. & S29\\  
2018-07-12& 0101.B-0570(A)& VLT, NACO, K-band & AO pos. & S29\\
2018-07-13 & 0101.B-0570(A) & VLT, NACO, H-, K-band & AO pos. & S2, S29, S38\\
2018-08-08 & 598.B-0043(O) & VLT, NACO, K-band & AO pos. & S2, S29\\
2018-08-09 & 598.B-0043(P)   & VLT, NACO, H-band & AO pos. & S2, S29, S38\\
2018-08-17 & 598.B-0043(O)  & VLT, NACO, K-band & AO pos. & S2\\
2019-04-19 & 5102.B-0086(C)  & VLT, NACO, K-band & AO pos. & S2, S29\\
2019-06-02 & 5102.B-0086(E)  & VLT, NACO, K-band & AO pos. & S2, S29\\
2019-07-03 & 5102.B-0086(F) & VLT, NACO, K-band & AO pos. & S2, S29\\
2019-09-01 & 5102.B-0086(H)  & VLT, NACO, K-band & AO pos. & S2, S29\\
2019-09-26 & 5102.B-0086(H) & VLT, NACO, K-band & AO pos. & S2, S29
 \end{tabular}
 }
 \end{center}
\end{table*}

\end{appendix}

\end{document}